\documentclass{elsart}
\usepackage{natbib}
\usepackage{epsfig}
\begin{document}
\runauthor{Vassiliev}
\begin{frontmatter}
\title{Extragalactic background light absorption signal
in the TeV $\gamma$-ray spectra of blazars.}
\author{V.\ V.\ Vassiliev\thanksref{corr_author}}

\address{Whipple Observatory, Harvard-Smithsonian CfA, 
P.O. Box 97, Amado, AZ 85645, USA}

\begin{abstract}
Recent observations of the TeV $\gamma$-ray spectra of the two closest
active galactic nuclei (AGNs), Markarian 501 (Mrk 501) and Markarian
421 (Mrk 421), by the Whipple and HEGRA collaborations have 
stimulated efforts to estimate or limit the spectral energy density (SED)
of extragalactic background light (EBL) which causes attenuation of
TeV photons via pair-production when they travel cosmological distances.
In spite of the lack of any distinct cutoff-like feature in the 
spectra of Mrk 501 and Mrk 421 (in the interval $0.26-10$ TeV) 
which could clearly indicate the presence of such a photon absorption 
mechanism, we demonstrate that strong EBL attenuation signal 
(survival probability of $10$ TeV photon $<10^{-2}$) may still be 
present in the spectra of these AGNs. This attenuation could escape 
detection due to ambiguity of spectra interpretation between
intrinsic properties of the sources and absorption by EBL.  
By estimating the minimal and maximal opacity of the universe to 
TeV $\gamma$-ray photons, we calculate the visibility range for current 
and future $\gamma$-ray observatories, and show that the Whipple 
$\gamma$-ray telescope should be able to detect (in $10$ hours 
at a $5\sigma$ confidence level) a BL Lac object with properties 
similar to Mrk 501 during its peak activity located at distances up 
to $z=0.12$. The proposed atmospheric Cherenkov telescope array VERITAS
should be able to see such an object at least as far as $z=0.3$.
Finally, we show that the proposed experiments, VERITAS, HESS, and MAGIC, 
may even be able to actually {\em measure} the EBL SED because their 
observations extend to the critical $75-150$ GeV regime. 
In this transition region a distinct ``knee-like'' feature should 
exist in the spectra of blazars, which is invariant with respect 
to their intrinsic properties. The change of the spectral index and flux 
amplitude across this knee, if observed for several blazars, will 
provide missing pieces of information needed to measure EBL in 
the wavelength range $0.1-30$ $\mu$m. 
\end{abstract}



\thanks[corr_author]{Corresponding author:
vvassiliev@cfa.harvard.edu}


\end{frontmatter}

\section{Introduction}

It was noted first by Gould and Schr\'{e}der \cite{GS67b} in the late 
sixties ``that observations of cosmic photons in the region 
$10^{12}$ to  $10^{13}$ eV would be of great value, since 
in this region absorption due to the cosmic optical photons is 
important. In fact, this may provide a means of determination 
of the optical photon density and of testing cosmological models.''
Although the authors pointed out that quasars may be the potential 
sources for such observations, and the ground-based Cherenkov technique 
could be the means of study, the real possibility to utilize this 
idea came one quarter of a century later with the discovery
of TeV extragalactic $\gamma$-ray sources by the Whipple 
collaboration. These are low-redshift BL Lac objects Markarian 421
$(z=0.031)$ \cite{Punch92}, Markarian 501 $(z=0.033)$ \cite{Quinn96},
and 1ES2344+514 $(z=0.044)$ \cite{Catanese98}. The detection of a more
distant X-ray BL Lac object, PKS 2155-304 $(z=0.117)$, has been 
recently reported by the Durham group \cite{Chadwick99}.

Interest in the absorption of extragalactic $\gamma$-radiation in 
recent years has been supported by two reasons: understanding 
the observational range for extragalactic sources of the very high 
energy photons (above $\sim 1$ TeV) and estimating the density of the 
EBL photons of rather low energy (below $\sim 1$ eV). The first motivation 
is of great importance for future $\gamma$-ray experiments, such as
VERITAS~\cite{VERITAS}, HESS~\cite{HESS}, and MAGIC~\cite{MAGIC}, 
and for testing of blazar emission models \cite{Dermer92, Marashi92, 
Mannheim93, Sikora94}. The second attracts great theoretical attention, since 
EBL in the energy range $\sim 10^{-2}-10$ eV contains a wealth of information 
on the early history of the universe providing integral constraints on 
galaxy and star formation, the initial distribution of stellar masses, 
metal and dust production, and the rate of re-processing of starlight 
to infrared wavelengths by the dust \cite{Dwek98, Primack99}. It is also 
possible that a significant fraction of the EBL is contributed 
by brown dwarfs, accreting black holes in active galactic nuclei 
(AGNs), and the decay of relic particles \cite{BCH86, BCH91}. 
Because the EBL contains a significant fraction of all energy released 
since the recombination epoch, its determination may set a limit on 
the energetic balance of the universe and constrain
particular energy sources \cite{Dwek98}.

Diffuse Infrared Background Experiment ({\em DIRBE}) and Far 
Infrared Absolute Spectrometer ({\em FIRAS}) instrument on board 
the Cosmic Background Explorer ({\em COBE}) spacecraft \cite{Boggess92} 
have been used to complete an effort to directly detect 
an isotropic EBL in ten photometric bands from $1.25$ to $240$ $\mu$m
\cite{Hauser98}, and $125-2000$ $\mu$m wavelength interval \cite{Fixsen98}.
The detection of the EBL for wavelengths shorter than $140$ $\mu$m
have been greatly impaired by the presence of strong backgrounds
from interplanetary dust scattering and emission, stars and galactic 
foregrounds, and from interstellar dust emission. Removing these 
various contaminating radiations to detect a directionally
isotropic signal, which is presumably a unique feature of EBL,    
is a difficult task \cite{Kelsall98, Arendt98}. The {\em COBE}
collaboration reported a detection of the EBL signal at $140$
and $240$ $\mu$m, but established only upper limits in the region
$1.25-100$ $\mu$m \cite{Hauser98}, the most important interval for 
attenuation of extragalactic TeV $\gamma$-rays. Thus, observation of 
TeV and sub-TeV extragalactic sources is currently the only relatively 
direct method which may provide new insight on the distribution of 
EBL in this wavelength range.

There are two alternatives which can be used to constrain the EBL
spectral energy density from $\gamma$-ray observations. First is to model 
EBL spectra based on the knowledge of galaxy and star evolution,
and use it to find the optical depth $\tau(E,z)$ and survival probability
$\exp(-\tau(E,z))$ for $\gamma$-rays of energy $E$ traveling 
from a source at redshift $z$. The spectrum of a particular
source can be interpreted then as modified by energy-dependent EBL 
attenuation. Such predictions of the opacity of the universe to 
$\sim$TeV photons have been reported in~\cite{MacMinnPrimac96, 
Primack99, SteckerDeJager98, StanevFranceschini98}. Many more
models have been suggested which derive the EBL SED based
on observational constraints and theoretical assumptions
(for a summary and references see~\cite{Dwek98}). These EBL predictions, 
which can be easily converted into a $\gamma$-ray optical depth, are roughly 
classified in two categories: backward and forward evolution models. 
The former models adopt current observations of spectra of galaxies as a 
function of luminosity and translate them backward in time using data 
on the redshift evolution of galaxy emissivity. An example of such 
empirically based calculations is given by 
Malkan \& Stecker~\cite{MalkanStecker98}. 
The latter models tend to simulate the evolution 
of galactic systems forward in time assuming initial parameters which 
do not conflict with current observational data, 
e.g.~\cite{SomervillePrimack98, Guiderdoni97}. A clear advantage of these
models is the possibility to test various scenarios, and, if the EBL is 
observed to have certain features in a spectrum, it may be possible to deduce 
information about cosmology, galaxy and star formation, as well as 
absorption and radiation of the light by dust. Both backward and 
forward evolution models do not guarantee, however, that all potential 
contributors to the EBL are accounted for and therefore more 
direct constraints on EBL spectral energy density are desirable.
Such an alternative, which attempts to unfold the EBL SED from
spectral measurements of extragalactic sources, has been explored 
recently by Stanev \& Franceschini, and Biller et 
al.~\cite{StanevFranceschini98, Biller98}. In this paper we 
follow an analogous strategy to establish limits on EBL from the spectral 
measurements of two AGNs, Mrk 421 and Mrk 501, and define 
applicability conditions of such an approach.

The interpretation of the EBL limits derived from TeV $\gamma$-ray
measurements is difficult because the EBL absorption cannot be completely 
isolated from the properties of the source. Although, this problem may 
be overcome in the future when more TeV and sub-TeV sources are discovered 
and the physics of the $\gamma$-ray emission from blazars is better 
understood, at present, this feature of the method remains a major source of 
uncertainty. The credibility of these derivations of the EBL limits from TeV 
AGN $\gamma$-ray measurements is still not accepted~\cite{Hauser98,Dwek98}. 
The claim of the first measurement of the EBL by de Jager et 
al.~\cite{DeJager94} derived from an apparent ``cutoff'' in the spectra 
of Mrk 421 at $4.1$ TeV~\cite{Mohanty93} has not been confirmed. 
More recent measurements of the spectra of Mrk 421 and 
Mrk 501~\cite{Krennrich97, Krennrich98, Konopelko99} are of much better
quality but the interpretation is not unambiguous. The Whipple
group~\cite{Krennrich98} find curvature in the spectrum of Mrk 501
but conclude that the spectrum of Mrk 421 is consistent with a 
simple power law up to 10 TeV. The curvature in the spectrum of
Mrk 501 is confirmed by the HEGRA group with high precision~\cite{Konopelko99}.
Because of the apparent differences in the curvature in the spectra
of the two AGNs (significant only at the $\sim 2$ sigma level), 
Krennrich et al.~\cite{Krennrich97} suggest there is no evidence
for EBL absorption. Stecker has recently revised an earlier 
prediction~\cite{SteckerDeJager98} and now suggests that the new 
spectrum of the flaring Mrk 501 shows EBL absorption and agrees with one 
of his EBL models~\cite{Stecker99}. The recent investigation by 
Stanev \& Franceschini of the Mrk 501 spectrum~\cite{StanevFranceschini98} 
has indicated upper limits on the EBL which almost come into conflict 
with observational evaluations based on deep surveys of extragalactic 
sources. These authors conclude that ``if spectra at TeV energies for 
extragalactic $\gamma$-ray sources like this for Mrk 501 are confirmed 
with improved statistics, we may be forced to conclude that the process of 
$\gamma-\gamma$ interaction in the intergalactic space is more complex 
than expected and that the average intergalactic magnetic field is 
extremely weak ($B<10^{-11}$ G).''

In this paper we strive to reconcile these differing interpretations, we 
attempt to limit the scope of the potential constraints which $\gamma$-ray 
observations may set on the SED of EBL, and we try to confirm some of the 
earlier results derived from TeV $\gamma$-ray measurements. The structure 
of the paper is as following. The next section, \S\ref{EBLsummary}, is a 
summary of the direct EBL experimental detections, upper and low limits. 
These are important for unfolding SED of EBL from the spectra
of blazars because they can be used to normalize the EBL contribution 
to $\gamma$-ray absorption and establish upper and low limits on 
TeV $\gamma$-ray survival probability. Then, in section \S\ref{EBLinteraction},
we consider details of the pair production process, which absorbs high energy 
photons, and point out several features important for understanding of this 
fundamental interaction. We also repeat why one would expect a cutoff
in the spectra of TeV extragalactic sources caused by the EBL. 
The next section, \S\ref{AGNspectra}, analyzes the recently reported spectra 
of two AGNs which provide data over a wide dynamic range, with high 
statistical quality. Although no apparent cutoffs have been detected up to 
$10$ TeV, both spectra are consistent with a non zero curvature. In section 
\S\ref{EBLunfolding} we demonstrate that although a cutoff feature is 
possible, it is not necessarily the only indicator of EBL absorption 
and the lack of a cutoff in the spectra of AGNs does not automatically 
mean low density of EBL. Such an alternative is explored in detail in 
this section, and we derive a SED of EBL which eludes detection due to
ambiguity in the interpretation of spectral measurements. Then we
determine upper and low limits of this EBL component. In section 
\S\ref{EBLattenuation}, we use these data to predict maximal and minimal 
attenuation in the spectra of a ``Mrk 501-like'' blazar at high redshift 
if such a hypothetical object demonstrates a peak of activity, and 
compare its flux with the sensitivities of the existing Whipple~\cite{Weekes89}
and proposed VERITAS~\cite{VERITAS} $\gamma$-ray observatories. 
The last section, \S\ref{Discussion}, is devoted to a discussion.

\section{Summary of direct EBL detections, upper and low limits}
\label{EBLsummary}

\begin{figure}
\centerline{\epsfig{file=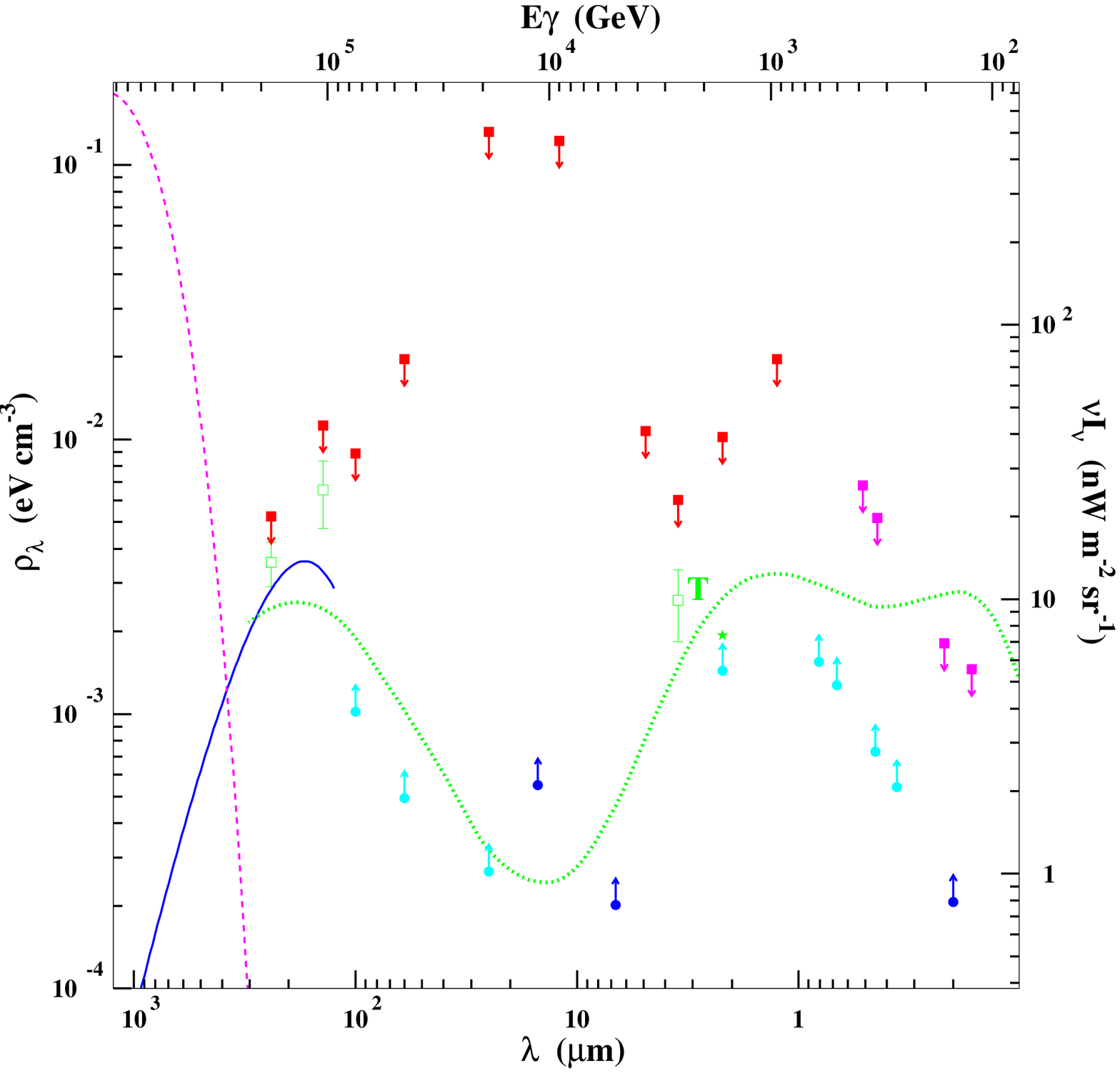,height=5.5in}}
\caption{Filled squares with arrows indicate EBL upper limits (95\%CL)
determined by various experiments. The wavelength range from 
$1.25$ $\mu$m to $240$ $\mu$m is from the DIRBE experiment~\cite{Hauser98}. 
Optical and UV bands are by all-sky-photometry observations~\cite{Bowyer91, 
Maucherat-Joubert80, Toller83, Dube79}. The open squares at 
$140$ $\mu$m and $240$ $\mu$m are DIRBE detections (shown with 
$1\sigma$ error bars). A tentative detection~\cite{DwekArendt98} 
is indicated by open square marked ``T''.
The filled circles indicate lower limits (95\%CL) 
obtained from galaxy counts. The range $25-100$ $\mu$m 
is from IRAS data~\cite{Hacking91}. The $6.7$ and $15$ 
$\mu$m are from ISO measurements~\cite{Oliver97}.
The ground-based measurement at $2.2$ $\mu$m is reported 
in~\cite{Pozzetti96}. The lower limits in the range from 
$0.36$ $\mu$m to $0.81$ $\mu$m are derived from HDF galaxy 
counts~\cite{Pozzetti98}. The UV lower limit at $0.2\mu$m 
is from~\cite{Armand94}. The solid line is a FIRAS detection 
of the far-infrared EBL~\cite{Fixsen98}. The dashed line 
indicates $2.7$K CMB radiation. The dotted line is 
an example of EBL prediction found by Primack et al. 
for the ``LCDM Salpeter'' model in~\cite{Primack99}.
\label{EBLdata}
}
\end{figure}

Fig.~\ref{EBLdata} depicts the current observational results 
on the EBL from UV to far-infrared wavelengths. The range 
from $1.25$ $\mu$m to $240$ $\mu$m is covered by DIRBE results 
obtained after a complex background subtraction~\cite{Hauser98}. 
Detections are found at $140$ $\mu$m and $240$ $\mu$m. Upper limits 
are reported for the other bands where foreground emission from the 
galaxy and interplanetary dust did not allow estimation of the EBL SED. 
A tentative detection has also been reported at $3.5$  
$\mu$m~\cite{DwekArendt98}, derived with the use of the DIRBE upper 
limit and a lower limit from ground-based K-band galaxy 
counts~\cite{Gardner96}. The latter, shown with the filled star 
in Fig.~\ref{EBLdata}, was assumed to be a detection in this analysis. 
The upper limits in UV and optical are derived from all-sky-photometry 
measurements~\cite{Bowyer91, Maucherat-Joubert80, Toller83, Dube79}. 
For a more extended review of earlier results in this energy 
range see~\cite{Cowie88}. Dwek et al.~\cite{Dwek98} also
indicate recent detections in this wavelength range by Bernstein 
et al. (not shown). 


The lower limits obtained from galaxy counts have been derived 
from IRAS data~\cite{Hacking91} for $25$, $60$, $100$ $\mu$m.
The two data points at $6.7$ and $15$ $\mu$m are deduced from 
ISO data~\cite{Oliver97} as reported in~\cite{StanevFranceschini98}. 
Pozzetti et al. published ground-based measurements in 
K-band~\cite{Pozzetti96}, and derived recently lower limits from 
the Hubble Deep Field (HDF) observations~\cite{Pozzetti98} covering the 
range $0.36-0.81$ $\mu$m. The ground-based UV lower limit at 
$0.2\mu$m is reported by Armand et al.~\cite{Armand94}. The solid line in 
Fig.~\ref{EBLdata} indicates a FIRAS detection of the 
extragalactic far-infrared background~\cite{Fixsen98} in the 
range $125-2000$ $\mu$m. The cosmic microwave background 
dominates EBL for wavelengths above $\sim 400$ $\mu$m.

The dotted line in Fig.~\ref{EBLdata} shows an example EBL 
prediction modeled by Primack et al.~\cite{Primack99} for a cold dark 
matter cosmology with non-zero cosmological constant
and Salpeter stellar initial mass distribution~\cite{Salpeter}. 
The main features of the EBL curve, common for the majority of models, are
a peak in SED distribution in the range from $0.1$ to $10$
$\mu$m formed mainly by starlight accumulated and redshifted 
through the history of the universe, the so-called stellar EBL component,
and a second peak in the range $10-1000$ $\mu$m formed by dust 
re-emission of the previously absorbed starlight, the dust EBL component. 
The excess of SED prediction in the UV region can be adjusted to 
observations in this model either by a higher starlight extinction 
rate by the dust, or by changing the stellar initial 
mass distribution to the Scalo model~\cite{Scalo}, which has less
mass concentrated in high-mass stars and consequently produces
less UV light. The EBL predictions by the different models vary,
and some of them generate less pronounced features, e.g.\
Malkan \& Stecker calculated an EBL with a substantially smaller 
valley in the $10$ $\mu$m region~\cite{MalkanStecker98}. 
This difference is discussed in the last section. 

\section{The absorption of extragalactic TeV photons}
\label{EBLinteraction}

\begin{figure}
\centerline{\epsfig{file=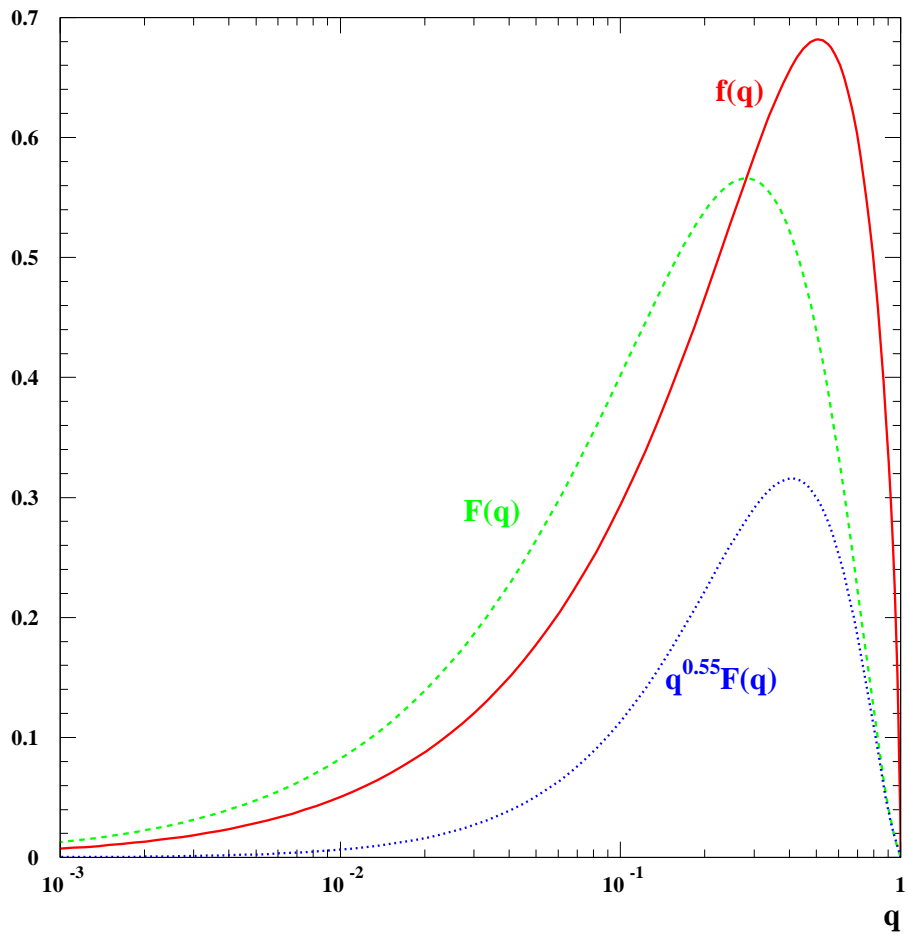,height=5.5in,angle=0.}}
\caption{ The behavior of the functions $f(q)$, $F(q)$, 
and $q^{0.55}F(q)$ which peak at $q=0.51$, $0.28$, and $0.41$ respectively.}
\label{functions}
\end{figure}

We are interested in the process $\gamma+\gamma \rightarrow e^+ + e^-$,
for which the total cross section is
\begin{eqnarray}
\sigma(q) & = & \frac{3}{8}\ \sigma_{\mathrm{T}} \ f(q) \nonumber \\ 
f(q)      & = & q \left[(1+q-\frac{q^2}{2})\ln 
                \frac{1+\sqrt{1-q}}{1-\sqrt{1-q}}
                -(1+q)\sqrt{1-q} \right] \label{cross-section} \\
q & = & \frac{m_{\mathrm{e}}^2}{E\varepsilon}\ 
\frac{2}{1-\cos(\theta)} \nonumber  ,
\end{eqnarray}
where $\sigma_{\mathrm{T}}=6.67 \times 10^{-25}$ cm$^2$ is Thomson scattering 
cross-section, $m_{\mathrm{e}}=0.51$ MeV is the mass of the electron, $E$ and 
$\varepsilon$ are the high and low energies of the photons, 
and $\theta$ is the collision angle (see, for example,~\cite{Heitler}). 
The $f(q)$ function, shown in Fig.~\ref{functions}, reaches its maximum 
at $q=0.508$, indicating that for a {\em head on} collision the peak of the 
interaction cross-section of the $\gamma$-ray photon of energy $E \sim 1$ 
TeV corresponds to pair production with a soft photon of energy 
$\varepsilon \sim 0.5$  eV ($\lambda \sim 2.5$ $\mu$m)
\begin{eqnarray}
\lambda &=& 2.41\mbox{$\mu$m}\ \frac{E}{\mbox{TeV}}.
\label{headonpeak}
\end{eqnarray}

The absorption probability of the high energy photon per unit path 
length $\d l$ by isotropic diffuse radiation with spectral density  
$\d n(\varepsilon)/\d \varepsilon$, is given by (see, for example, 
original calculations by Gould \& Schr\'{e}der \cite{GS67a})
\begin{eqnarray}
\frac{\d \tau}{\d l} & = & \frac{3}{8}\ \sigma_{\mathrm{T}} \ 
\int \d \varepsilon \ 
\int_{\frac{m_{\mathrm{e}}^2}{E\varepsilon}}^{1} \ \d x \ 
f(\frac{m_{\mathrm{e}}^2}{E\varepsilon}\ \frac{1}{x})\ 
2x\ \frac{\d n(\varepsilon)}{\d \varepsilon} , \label{absorbP} \\
x & = & \frac{1-\cos \theta}{2}. \nonumber
\end{eqnarray}
Eq.\ (\ref{absorbP}) can be transformed to the following form
\begin{eqnarray}
\frac{\d \tau}{\d l} & = & \frac{3}{8}\ \sigma_{\mathrm{T}} \ 
\int_{\frac{m_{\mathrm{e}}^2}{E}}^{\infty} \d \varepsilon \ 
\frac{\d n(\varepsilon)}{\d \varepsilon}\ 
F(\frac{m_{\mathrm{e}}^2}{E\varepsilon}), \label{absorbF} \\
F(q) & = & 2\ q^2\ \int_{q}^{1} f(x)x^{-3} \d x. \label{F(q)}
\end{eqnarray}
Therefore, attenuation of the high energy $\gamma$-rays by {\em isotropic}
background photons peaks at the maximum of $F(q)$ 
($q=0.28)$, shown in Fig.~\ref{functions}. This gives
a characteristic energy of the soft photon $\varepsilon \sim 0.9$ eV
for the most effective attenuation of a $1$ TeV $\gamma$-ray
\begin{eqnarray}
\lambda &=& 1.33\mbox{$\mu$m}\ \frac{E}{\mbox{TeV}}.
\label{isotropicpeak}
\end{eqnarray}
This energy is almost a factor of two higher than for a head on 
collision given in Eq.\ (\ref{headonpeak}). The upper axis in 
Fig.~\ref{EBLdata} indicates energies of the $\gamma$-ray photons
coupled to the wavelengths of the EBL photons 
via Eq.\ (\ref{isotropicpeak}).

The lowest energy of a soft photon important for interaction
with a hard photon of energy $E$ is determined by the threshold for
pair production in a head on collision
\begin{eqnarray}
\lambda &=& 4.75\mbox{$\mu$m}\ \frac{E}{\mbox{TeV}}.
\label{tresholdenergy}
\end{eqnarray}
Thus, for example, a $100$ TeV $\gamma$-ray begins to feel the 
presence of the CMB, but it is attenuated mostly by the dust EBL 
component. 


The highest energy $\varepsilon$ relevant to interactions with 
extragalactic photons of energy $E$ should be derived from the 
properties of the function $F(q)$. The asymptotic behavior of this function
at small $q$ (large $\varepsilon$)
\begin{eqnarray}
\lim_{\varepsilon \rightarrow \infty} 
F(\frac{m_{\mathrm{e}}^2}{E\varepsilon}) & \rightarrow &
2 \frac{m_{\mathrm{e}}^2}{E\varepsilon} 
(\ln \frac{4 E \varepsilon}{m_{\mathrm{e}}^2}-2)
\label{assymptotic}
\end{eqnarray}
converges very slowly to zero, such that the integral 
\begin{eqnarray}
\int_{\frac{m_{\mathrm{e}}^2}{E}}^{\infty} 
\d \varepsilon \ F(\frac{m_{\mathrm{e}}^2}{E\varepsilon}) 
& \rightarrow & \infty \nonumber
\end{eqnarray}
is divergent at the upper limit. This indicates that interaction 
of a hard photon of energy $E$ extends far below the 
characteristic wavelength given by Eq.\ (\ref{isotropicpeak}), 
and suppression of the integral (\ref{absorbF}) divergence at 
large $\varepsilon$ is due to the behavior 
of the EBL spectral density $\d n(\varepsilon)/\d \varepsilon$.
It follows then, that the SED of EBL itself defines the characteristic 
energy of the soft photon and range of the photon's energy important 
for intergalactic absorption of a $\gamma$-ray with energy $E$.

As an important example we consider a SED of the form
\begin{eqnarray}
\frac{\d n(\varepsilon)}{\d \varepsilon} &=&
\frac{\alpha}{\mbox{eV}}
\left(\frac{\varepsilon}{\mbox{eV}} \right)^{-\beta},
\label{spdn}
\end{eqnarray}
and derive the differential absorption probability
\begin{eqnarray}
\frac{\d \tau}{\d l}&=&\frac{3}{8}\ \sigma_{\mathrm{T}} \ \alpha \ g(\beta) 
\left(\frac{E}{\mbox{TeV}} \right)^{\beta-1}, \label{powerlaw} \\
g(\beta)&=& \left(\frac{\mbox{MeV}}{m_{\mathrm{e}}}\right)^{2(\beta-1)}
\int_0^1 q^{\beta-2}\ F(q)\ dq. \nonumber
\end{eqnarray}
Depending on the power index $\beta$, the main contribution to this 
integral comes from the different regions $q$, and, therefore, from
different energy ranges of the EBL field. For $\beta=2.55$ the function 
$q^{0.55}F(q)$ is shown in Fig.~\ref{functions}. It peaks at $q=0.41$ and 90\%
of the contribution to the integral $g(2.55)=1.47$ is accumulated when 
$q$ changes from $0.13$ to $0.80$. Only $5$ percent tails are left
below and above this interval. Thus, the range of soft photon 
wavelengths important for attenuation of $\gamma$-ray of energy $E$ is
\begin{eqnarray}
0.62 \mbox{$\mu$m}\ \frac{E}{\mbox{TeV}}<\lambda
<3.8 \mbox{$\mu$m}\ \frac{E}{\mbox{TeV}}. \label{r255}
\end{eqnarray}
The similar limits derived for $\beta=2.00$, when value of $g(2.00)=1.25$, are
\begin{eqnarray}
0.37 \mbox{$\mu$m}\ \frac{E}{\mbox{TeV}}<\lambda
<3.6 \mbox{$\mu$m}\ \frac{E}{\mbox{TeV}}. \label{r200}
\end{eqnarray}
It can be seen that while low energy end of soft photons remains
almost the same for these two examples, the high energy end changes 
by a factor $~1.7$ extending the range important for attenuation 
of $\gamma$-ray photons to high energies in the case $\beta=2.00$.

For a $\gamma$-ray of energy, $E$, which travels cosmological distances 
from a source at redshift $z$ we generalize Eq.\ (\ref{absorbF}) 
following, for instance~\cite{Stecker71}, to obtain the optical depth
\begin{eqnarray}
\tau(E,z) & = & \frac{1}{n_0\ h_0} 
\int_0^z \sqrt{1+z}\ \d z \
\int_{\frac{m_{\mathrm{e}}^2}{E (1+z)^2}}^{\infty}
\d \varepsilon \ 
\frac{\d n(\varepsilon)}{\d \varepsilon}\ 
F(\frac{m_{\mathrm{e}}^2}{E \varepsilon (1+z)^2}), \label{OpticalDepth} \\
\frac{1}{n_0\ h_0}& = & \frac{3}{8}\ \sigma_{\mathrm{T}} \ \frac{c}{H_0},
\nonumber 
\end{eqnarray}
where $H_0=100\ h_0$ km s$^{-1}$ Mpc$^{-1}$ is the Hubble 
constant, $0.5<h_0<0.85$ is the normalized Hubble expansion rate,
$c$ is the speed of light, $n_0=4.33 \times 10^{-4}$ cm$^{-3}$
is the characteristic density, and $\d n(\varepsilon)/\d \varepsilon$ is 
the present-day spectral EBL density. It is also assumed that the 
density parameter of the universe $\Omega=1$, cosmological constant
$\Omega_{\Lambda}=0$, and evolutionary considerations of the EBL 
spectral density are neglected, which is justified for ($z<0.3$), 
implying that the EBL has been produced mostly at higher 
redshifts~\cite{Madau96}. 

The appearance of a cutoff in the spectra of TeV extragalactic 
sources has been suggested in one of the pioneering works 
on TeV $\gamma$-ray absorption by Stecker, de Jager and 
Salamon~\cite{SDS92} for quasar 3C 279 ($z=0.54$) and 
later several predictions~\cite{SD93, DeJager94, DwekSlavin94, Biller95} 
for Mrk 421 have been published.  If the differential
energy spectrum of a source obeys a power-law,
characterized by index $\delta$ and flux $J$ one
would expect to observe a spectrum modified by EBL absorption
\begin{eqnarray}
\frac{\d N}{\d E} & = & J\ E^{-\delta}\exp (-\tau(E,z)).
\label{spectrum}
\end{eqnarray}
For an EBL spectral density (\ref{spdn}) and small $z$, the
optical depth is given by
\begin{eqnarray}
\tau(E,z) & = & \frac{\alpha\ g(\beta)\ z }{n_0\ h_0}
\left(\frac{E}{\mbox{TeV}} \right)^{\beta-1}
=\left(\frac{E}{E_{\mathrm{c}}} \right)^{\beta-1},
\label{PLopticaldepth} \\
E_{\mathrm{c}} & = &
\left(\frac{n_0\ h_0} {\alpha\ g(\beta)\ z } \right)^{\frac{1}{\beta-1}}
\mbox{ TeV}, \nonumber
\end{eqnarray}
and the prediction for an observed spectrum of the source is then
\begin{eqnarray}
\frac{\d N}{\d E}=J\ E_{\mathrm{c}}^{-\delta}
\exp\left(-\delta \ln \left(\frac{E}{E_{\mathrm{c}}} \right)
-\left( \frac{E}{E_{\mathrm{c}}} \right)^{\beta-1}\right).
\label{cutoffspectrum}
\end{eqnarray}
If $\beta$ is large (as in the original paper which 
claimed possible EBL measurement~\cite{DeJager94} from TeV
$\gamma$-ray observations: $\beta=2.55$, $\delta=2.1$), a sharp 
cutoff must exist in the spectrum because the power-law term in the 
exponent dominates the logarithmic term at energies above 
$E_{\mathrm{c}}$. A lack of events in the preliminary spectrum of 
Mrk 421 in bins above $3.4$ TeV and $5.1$ TeV was interpreted as 
such a cutoff providing an estimate of $E_{\mathrm{c}}=4.1$ TeV. 
Due to a rapid relative change of the two terms in the exponent, 
the logarithmic term dominates for energies below 
$0.73\ E_{\mathrm{c}}=3.0$ TeV leaving almost no trace of the 
cutoff at lower energies ($<0.3\ E_{\mathrm{c}}$). 
Although this detection has not been confirmed in later observations 
(see next sections \S\ref{AGNspectra}) the approach to place limits on
EBL by ``non-detection'' of the cutoff feature in AGN spectra remains 
legitimate. If the spectral measurements of the source extend without a 
cutoff to an energy $E_c$, the EBL spectral 
density (\ref{spdn}) is constrained by 
\begin{eqnarray}
\alpha & < & n_0 \frac{h_0}{g(\beta)\ z }
\left(\frac{\mbox{TeV}}{E_c} \right)^{\beta-1}.
\label{EBLcutoffL}
\end{eqnarray}
It is important to note that as $E_c$ changes in observations,
both the $\alpha$ limit and the ``window'' of EBL photons, which did not
produce a cutoff feature at $E_c$, change according 
to Eqs.\ (\ref{EBLcutoffL}, \ref{r255}, \ref{r200}). Because this limit
depends on the EBL SED model, the model itself must be justified by
current theoretical considerations in the corresponding 
wavelength interval. For example, the region $10-100$ $\mu$m is
believed to be approximated well with a SED of the form (\ref{spdn})
with $\beta$ in the interval $2-3$. Such approximation, however, is 
invalid for the $1-10$ $\mu$m wavelength range. Due to the sensitivity of the
$\gamma$-ray absorption rate to a rather extended interval of
EBL photons (properly explained in Eq.\ (\ref{assymptotic})),
the optical depth $\tau(E,z)$ is not particularly sensitive to 
small details of the EBL distribution, they are smoothed out by the 
integration (Eq.~\ref{OpticalDepth}). On the other hand, the absorption 
effect on $\gamma$-rays may be exponential, at least for some 
EBL models (Eq.~\ref{cutoffspectrum}); this is the feature which 
allows one to establish EBL {\em upper limits} by a non-detection of 
the cutoff. Recent Mrk 501 observations by the HEGRA 
collaboration~\cite{Konopelko99} seem to indicate a rapid 
decline in the spectrum of this AGN above $10$ TeV. 
If the systematics of HEGRA observations in this energy regime
are well understood and these measurements are confirmed,
an interpretation of this decline in the Mrk 501 spectrum 
by EBL absorption may constrain the spectral density in 
the $10-70$ $\mu$m range.

\section{Mrk 421 and Mrk 501 spectra}
\label{AGNspectra}

Krennrich et al.~\cite{Krennrich98} have recently published
the compiled TeV spectral data of two AGNs, Mrk 421 $(z=0.031)$ 
and Mrk 501 $(z=0.033)$, shown in Fig.~\ref{Mrk421Mrk501}. 
The high statistical quality of these measurements and wide
energy range, covering $0.26-10$ TeV, allowed for the first time 
detection of curvature in the Mrk 501 spectrum during its high state of 
activity. This variable blazar shows changes in the TeV $\gamma$-ray flux 
on a time-scale of several hours~\cite{Quinn98} with 
no evidence for temporal variability of the spectral shape 
observed by the Whipple~\cite{Samuelson98} and HEGRA~\cite{Aharonian99} 
collaborations. The spectrum remains statistically invariant 
from the absolute flux which may increase up to several 
Crab units\footnote{The Crab unit is defined as the $\gamma$-ray 
flux above $300$ GeV from the Crab Nebula~\cite{Hillas98},
i.e. $N_\gamma(E>300\;GeV) \approx 10^{-6}$ m$^{-2}$s$^{-1}$.}
during flaring activity~\cite{Quinn98, Aharonian99}. 


Mrk 421 is also highly variable showing bursts of radiation 
on a sub-hour time scale~\cite{Gaidos96}, with flux variations of up 
to $10$ Crab units~\cite{Zweerink97}. The spectrum of this 
blazar was derived from three data sets with average fluxes of 
$2.8$, $3.3$, and $7.4$ Crab units. Neither of these
sets has revealed any statistical differences in the 
spectrum shape justifying compilation of the data~\cite{Krennrich99b}.

Both spectra are rather featureless in the energy range $0.26-10$ TeV
and do not show an apparent cutoff which could be interpreted as 
an EBL absorption. Both spectra are well described by a flux dependence
\begin{eqnarray}
\frac{\d N}{\d E} & = &J\ 
\exp\left(-\delta \ln(E/\mbox{TeV}) - c  \ln^2(E/\mbox{TeV}) \right),
\label{curvedspectrum}
\end{eqnarray}
where $\delta$ is a spectral power index, and $c$ 
is a curvature. The best fit of such a representation 
of the data is shown in Fig.~\ref{Mrk421Mrk501}. It can be seen 
from the figure that the spectra differ; a $\chi^2$ test places the 
chance probability that they arise from the same parent distribution 
as $4\times 10^{-3}$~\cite{Krennrich98}. The differences in the
two spectra must be due to differences in the intrinsic emission 
spectra since they are at practically the same redshift; however
the fraction of curvature that is common to both spectra could 
be explained by EBL absorption.

\begin{figure}
\centerline{\epsfig{file=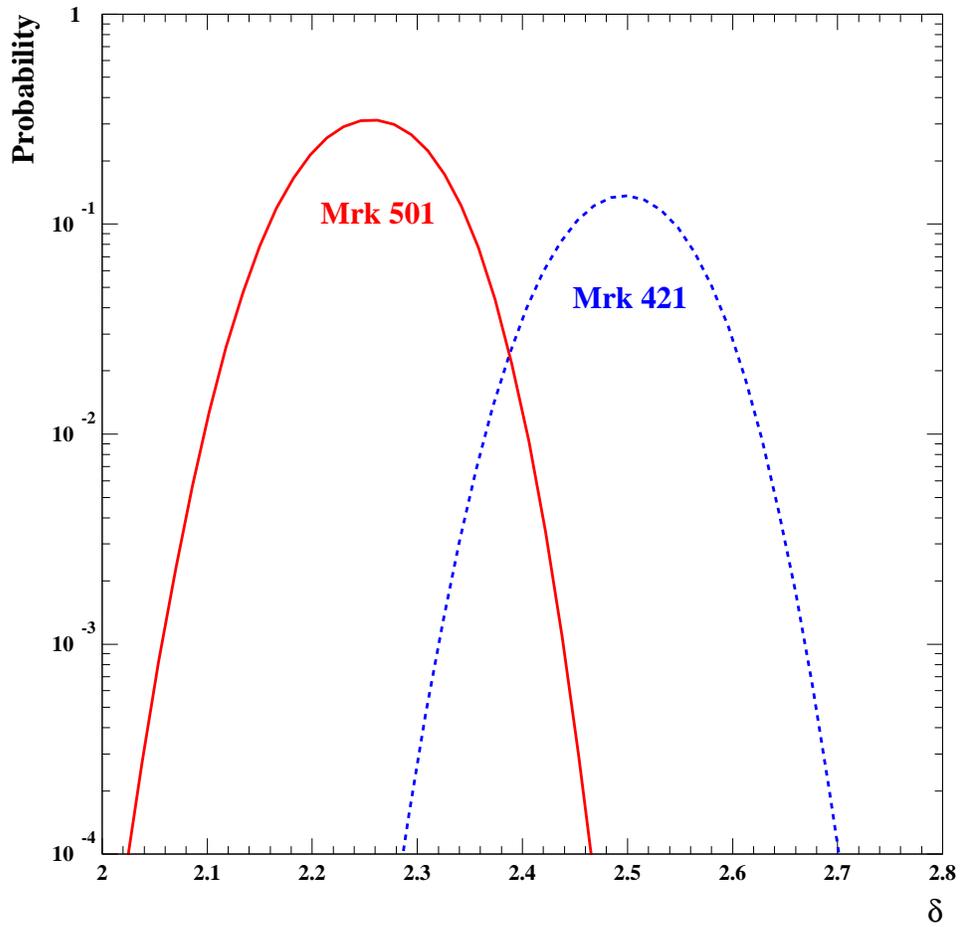,height=5.5in,angle=0.}}
\caption{The chance probability for AGN spectrum to have 
a certain value of spectral index $\delta$. Mrk 421 peak is 
at $\delta=2.50$ with $p_{\chi^2}\le 10^{-3}$ for $\delta \ge 2.68$, 
and Mrk 501 peak is at $\delta=2.26$ with $p_{\chi^2}\le 10^{-3}$ 
for $\delta \ge 2.44$.}
\label{index}
\end{figure}
\begin{figure}
\centerline{\epsfig{file=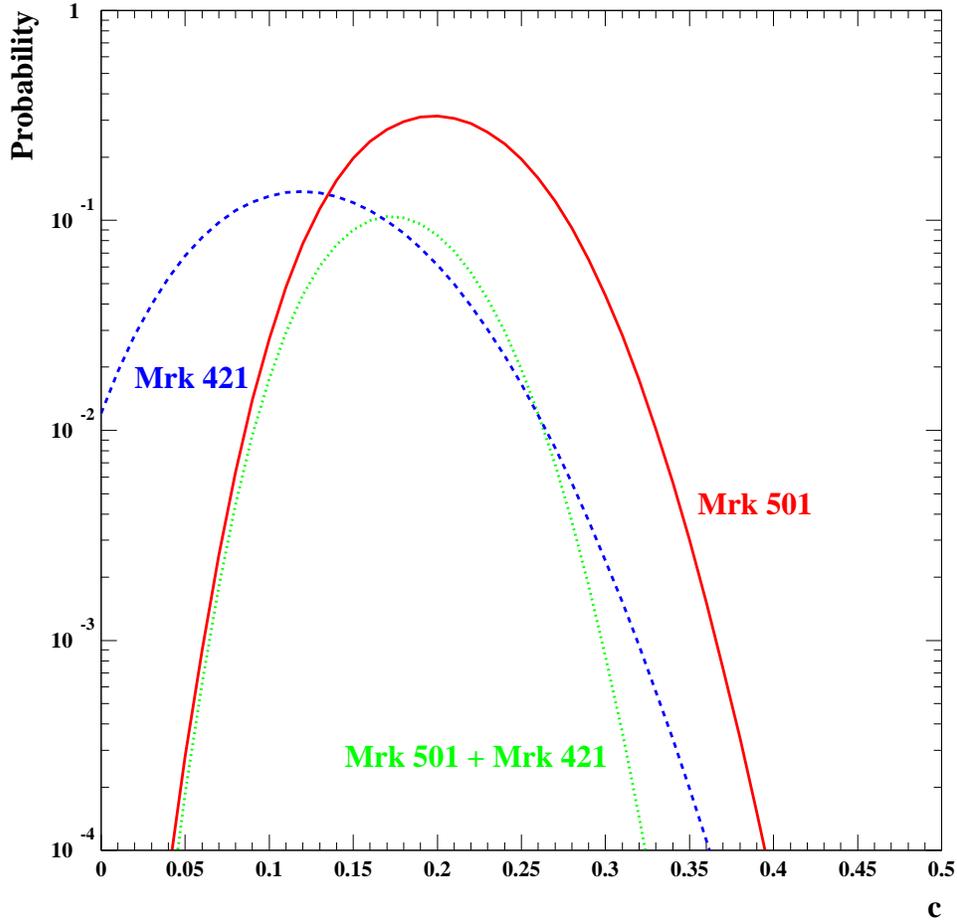,height=5.5in,angle=0.}}
\caption{ The chance probability for the AGN spectra to have 
a certain value of curvature parameter, $c$. Mrk 421 peak is 
at $c=0.12$ with $p_{\chi^2}\le 10^{-3}$ for $c \ge 0.32$, 
and Mrk 501 peak is at $c=0.20$ with $p_{\chi^2}\le 10^{-3}$ 
for $c \ge 0.37$. If curvature of the spectra is fixed for both 
AGNs but the spectral indices are allowed to vary, 
the combined chance probability for Mrk 421 and Mrk 501 is 
given by the dotted line. It peaks at $c=0.17$ with $(p_{\chi^2}=0.1)$,
Mrk 421 $\delta=2.48$, and Mrk 501 $\delta=2.28$. The chance that
$c \ge 0.30$ is less than $10^{-3}$.}
\label{curvature}
\end{figure}

The differences can arise from differences in the spectral index, $\delta$;
in the curvature parameter, $c$, or in a combination of both. The
simplest hypothesis is that a spectrum can be fit by a simple power law
($c=0$) and, as discussed in~\cite{Krennrich98}, the spectrum of
Mrk 421 is compatible with this hypothesis. The spectrum of Mrk 501 is not.
The spectral indices of the two blazars are statistically different.
Fig.~\ref{index} depicts the chance probability that a certain
value of spectral index can be associated with the data on each object.
The curvature parameter in these $\chi^2$ estimates has been 
fit to the best value at a given index. The statistical quality 
of the data is enough to rule out the hypothesis of equal spectral indices 
for the two AGN. A similar analysis of the curvature parameter is not 
definitive. Although the best fits to $c$ for each source are different
(Mrk 421 $0.12 \pm 0.04$; Mrk 501 $0.20 \pm 0.03$~\cite{Krennrich98})
it is also possible to fit both spectra with the same curvature 
parameter, Fig.~\ref{curvature}. Some or all of this curvature could 
arise from EBL absorption. The peak of the probability $(p_{\chi^2}=0.1)$ 
is at $c=0.17$ with the chance probability less than $10^{-3}$ that the 
curvature parameter exceeds $0.3$.


In the Whipple paper~\cite{Krennrich98} the differences in the spectra of 
two AGNs were interpreted in the context of synchrotron-inverse
Compton (SIC) blazar emission model, and it was 
concluded that the harder spectral index of Mrk 501 and its 
non-zero curvature is not inconsistent with such a model,
if X-ray observations of these objects are attributed to
synchrotron emission. It was pointed out that the difference
in the energy spectra of Mrk 421 and Mrk 501 
must be intrinsic to the sources and not due to 
intergalactic EBL absorption since both AGNs are located 
at the same distance $z=0.03$. It is shown in Fig.~\ref{curvature} 
that a substantial part of the curvature is not ruled out as common 
for both blazars, and therefore difference in their curvatures 
is to be considered as tentative only. Moreover, explaining the curvature of 
Mrk 421 and Mrk 501 as dominated by EBL absorption seems plausible in 
the context of SIC model because it is easier to understand then
the observed spectral shape independence of these blazars relative to their 
absolute flux during episodes of activity. When the intrinsic spectrum 
is almost a pure power-law, it remains invariant with respect to 
both amplitude and energy re-normalization. If a substantial curvature 
is present in the intrinsic spectrum, the latter invariance 
is lost, severely constraining models of blazar flaring 
(see for example~\cite{Mastichiadis97}).
If, however, the curvature of Mrk 501 is mostly due to 
EBL absorption, which is not inconsistent with observations 
of Mrk 421, then changes in the intrinsic hardness ratio 
of the AGN during flaring may be partially masked by this effect.  

In the recent submission~\cite{Djannati-Atai99} by the CAT 
collaboration on the spectral properties of Mrk 501 the Whipple result
on spectral index and curvature of this object has also been 
confirmed ($2.24$, $0.22$). Unlike the Whipple and HEGRA observations,
CAT data seem to indicate spectral variability depending on 
the state of emission. The detected change of hardness ratio is 
interpreted in the frame of a SIC blazar emission model with
synchrotron target photons, and an interpretation of Mrk 501 flaring 
activity is suggested as being due to injection of a population of
energetic electrons and/or an increase in acceleration efficiency.
Although the non-detection of spectral changes by the HEGRA
and Whipple collaborations is explained by high energy threshold 
and low statistics respectively, we think that this result needs
further confirmation to reliably disentangle the dynamic and static parts
of the spectral index and curvature parameters. The latter should 
presumably provide a better constraint on the EBL absorption effect.

Both the HEGRA and CAT collaborations have detected a signal above $10$
TeV in the spectrum of Mrk 501 during its highstate
activity~\cite{Aharonian99a, Djannati-Atai99} in 1997. The 
spectral measurements performed with the HEGRA stereoscopic system 
indicated the presence of an exponential cutoff at $6.2$ TeV in the 
power-law spectra of this object.  In the region $0.5-10$ TeV, however, 
the reported spectrum is consistent with the Whipple result and can be 
well described by the power-law with a curvature term. The region above $10$ 
TeV does not follow this dependence and shows a substantially more rapid 
decline in the AGN flux. Although the cutoff feature on its own
does not allow a unique interpretation in terms of EBL limits
and/or physical processes generating radiation at the source, 
it provides nevertheless a set of interesting constraints suggested
in~\cite{Aharonian99a}, among which an increase in the SED of EBL
in the region $30-70$ $\mu$m is a possibility. 

Because the cutoff above $10$ TeV in the spectra of AGNs has been 
observed only for one object and only by one instrument we do not 
discuss this feature in our paper due to a variety of alternatives in 
its explanation. Instead, we refer the reader to several recent
publications~\cite{Konopelko99a, Coppi99, Coppi99a}. Alternatively,
the region of the Mrk 501 spectrum below 10 TeV is well established
and confirmed by several groups including Whipple, CAT, HEGRA, the Telescope 
Array~\cite{Hayashida98}, and the Crimean observatory~\cite{Fomin99}.
In addition, spectral measurements of another AGN at a similar redshift, 
Mrk 421, are known in the same 
energy range and both spectra a well fitted by a power-law with 
the curvature term. The HEGRA collaboration has reported recently a Mrk 421 
spectrum in a low state of its activity~\cite{Aharonian99b} during 1997-1998.
Likely due to limited photon statistics, no nightly variability has been
found in the spectra of this blazar. The time averaged energy spectrum
has been found to be very steep with an index of $3.09 \pm 0.07$ in the 
energy range $0.5-7$ TeV. Although the significance of the difference 
between Whipple and HEGRA measurements and its impact on flaring spectral 
variability of this source remain to be understood, in both measurements the
time averaged AGN spectrum is consistent with a pure power-law, but an
exponential cutoff or curvature term can be equally well 
fitted to the data. We use Whipple data which spans the same energy interval  
($0.26-10$ TeV) to conservatively conclude that curvature, $c$, of the 
spectra of the two AGNs caused by intergalactic absorption does 
not exceed $0.3$. The curvature of both AGNs may be due to a
common cause, if so, its most probable value is $\sim 0.17$.

\section{EBL unfolding}
\label{EBLunfolding}

The detailed investigation of the spectra of two AGNs has not
revealed any ``sharp'' feature. The non-detection of the cutoff, however, 
does not automatically mean low EBL spectral density and negligible
intergalactic absorption of TeV $\gamma$-rays. The featureless 
spectra of blazars, in fact, allows a non-trivial interpretation.
The underlying reason for this is that EBL absorption
is difficult to separate from the properties of the source.
It can be seen from Eq.\ (\ref{cutoffspectrum}) that if parameter
$\beta$ of the EBL SED is close to $1$, the intergalactic absorption
does not produce a distinct feature in a source spectrum.  
We investigate such a possibility in this section. Our strategy
will be just opposite to a prediction of optical depth from EBL
spectral density. Instead we accept that $\tau(E,z)$ can be 
approximated by
 \begin{eqnarray}
\tau(E,z) & = & 
c_0(z) + c_1(z) \ln \left( \frac{E}{\mbox{TeV}} \right) 
+ c_2(z) \ln^2\left(\frac{E}{\mbox{TeV}} \right),
\label{Aopticaldepth}
\end{eqnarray}
which is supported by observations in the energy interval 
$(0.26-10)$ TeV for a small $z$. One can think of such an
approximation as an expansion series of the 
$\ln(E/\mbox{TeV})$ variable, when higher orders produce
undetectable contributions in a given energy interval for 
known spectra of blazars. Using this optical depth 
representation, we unfold the corresponding EBL spectral density.

To simplify our considerations, we assume at the moment that 
the redshift of the source is very small ($z \ll 1$). 
Using formula (\ref{OpticalDepth}) we obtain an integral equation
\begin{eqnarray}
\int_{\frac{m_{\mathrm{e}}^2}{E}}^{\infty} 
\d \varepsilon \ \frac{\d n(\varepsilon)}{\d \varepsilon}\ 
F(\frac{m_{\mathrm{e}}^2}{E\varepsilon})
& = & \frac{n_0\ h_0}{z}\ \tau(E), 
\label{Unfold1}
\end{eqnarray}
which can be resolved with the use of a {\em Mellin} transformation.
Omitting this derivation, we give the final result
\begin{eqnarray}
\varepsilon \ \frac{\d n(\varepsilon)}{\d \varepsilon} 
& = &\frac{n_0\ h_0}{z} \int_{\frac{m_{\mathrm{e}}^2}{\varepsilon}}^{\infty} 
\ K(\frac{m_{\mathrm{e}}^2}{E\varepsilon})\ \tau(E) \ \frac{\d E}{E}, 
\label{Unfold2} \\
K(q) & = & \frac{1}{2 \pi i} 
\int_{\sigma-i\infty}^{\sigma+i\infty} \frac{q^s\ \d s}{G(s)},  
\ \ \ \ (\sigma>-1), \nonumber \\
G(s) & = & \int_0^1\ q^{s-1}\ F(q)\ \d q. \nonumber 
\end{eqnarray}
It can be seen from Eq.\ (\ref{Unfold2}) that the formal 
solution $\d n(\varepsilon)/\d \varepsilon$ which satisfies the optical 
depth (\ref{Aopticaldepth}) may be represented as 
\begin{eqnarray}
\frac{\d n(\varepsilon)}{\d \varepsilon}
& = & \frac{n_0}{\varepsilon} \left(  
p_0-p_1 \ln \frac{\varepsilon}{\varepsilon_0}
+p_2 \ln^2 \frac{\varepsilon}{\varepsilon_0}
\right)
\label{EBLsolution}
\end{eqnarray}
with characteristic energy $\varepsilon_0$ and some constants 
$p_0, p_1, p_2$. We use this form of the EBL spectral density 
to connect $(p_0, p_1, p_2)$ to $(c_0,c_1,c_2)$ 
of the optical depth, conducting a direct calculation 
of integral (\ref{OpticalDepth}) with arbitrary, but not large 
redshift $(z<0.3)$ required by an applicability condition of this 
formula. One can find that
\begin{eqnarray}
c_i & = & \frac{1}{h_0} \sum_{j=1}^{3} p_j\ t_{ji}(z),
\label{CPconnection}
\end{eqnarray}
where $(i=1,2,3)$, and $t_{ji}(z)$ is a redshift dependent matrix.
The introduction of a finite redshift, $z$, does not 
change our conclusion about the form of the EBL SED (Eq.~\ref{EBLsolution})
since the integration order in integral (\ref{OpticalDepth}) can be 
changed, and $z$ should be considered then as a parameter. 
We also use a characteristic energy $\varepsilon_0=9.3$ eV
($\lambda=0.13$ $\mu$m) of the isotropic EBL which corresponds to a peak 
interaction with $100$ GeV $\gamma$-rays and truncate the SED at this energy.
The current TeV and sub-TeV measurements are not very sensitive to the EBL 
distribution in UV region. The truncation, however, will introduce an 
error in the optical depth for low energy $\gamma$-rays ($E\sim 100$ 
GeV) and larger $z$, limiting the applicability of our considerations 
to the $0.1-10$ TeV energy regime. We expect, though, that attenuation of 
$100$ GeV $\gamma$-rays will be rather small for the range of 
redshifts investigated here.

Three examples of EBL spectral density described by Eq.\ (\ref{EBLsolution}) 
are shown in Fig.~\ref{SEDlimits}. It is a curious coincidence that these
three parametric SED of the EBL can describe well the expected peak
in the $0.1-10$ $\mu$m wavelength range of the EBL energy density produced 
by starlight emitted and redshifted during evolution of the universe. 
There is no {\em a priori} reason why the current observational window 
of TeV astronomy ($0.26-10$ TeV) and requirement that EBL absorption 
of $\gamma$-rays must produce almost power-law modifications to the 
spectra of extragalactic sources would necessarily require that the
EBL SED, which is allowed to have a peak in the wavelength range 
$0.1-10$ $\mu$m, would coincide with the theoretical 
expectation based on completely different assumptions. 
In other words, if the starlight peak were to be shifted to a longer
wavelength region by a factor of $5-10$, any reasonable amount of 
EBL would necessarily produce a ``knee-like'' feature in the  
($0.26-10$ TeV) spectra of the sub-TeV sources. The starlight EBL peak 
is located in such a wavelength region that it eludes its apparent detection 
in the TeV and sub-TeV spectra of blazars due to ambiguity in the 
interpretation of the spectral measurements. 

   
The direct upper and low limits on the EBL SED (see Fig.~\ref{EBLdata})
restrict a possible choice of parameters $p_0$, $p_1$, and $p_2$.
For example, $p_0$, which defines the SED of EBL at $\lambda=0.13$ $\mu$m,
is likely limited to $(0.025-0.45)$ range by Armand et al.~\cite{Armand94}
$0.2$ $\mu$m lower and by Bowyer~\cite{Bowyer91} $0.165$ $\mu$m upper limits. 
While the exact choice of this parameter affects the optical depth of 
$\gamma$-rays with energies around $100$ GeV at large $z$, it has a rather 
negligible effect on the attenuation of photons in the $0.3-10$ TeV energy 
range. Thus its value is not constrained by current TeV observations of the 
low redshift sources. With the reservation that we do not predict correctly a 
small optical depth of $\sim 100$ GeV photons we fix this parameter somewhat 
arbitrarily, at $0.025$, assuming that EBL attenuation will unlikely be 
lower than the one we find.

\begin{figure}
\centerline{\epsfig{file=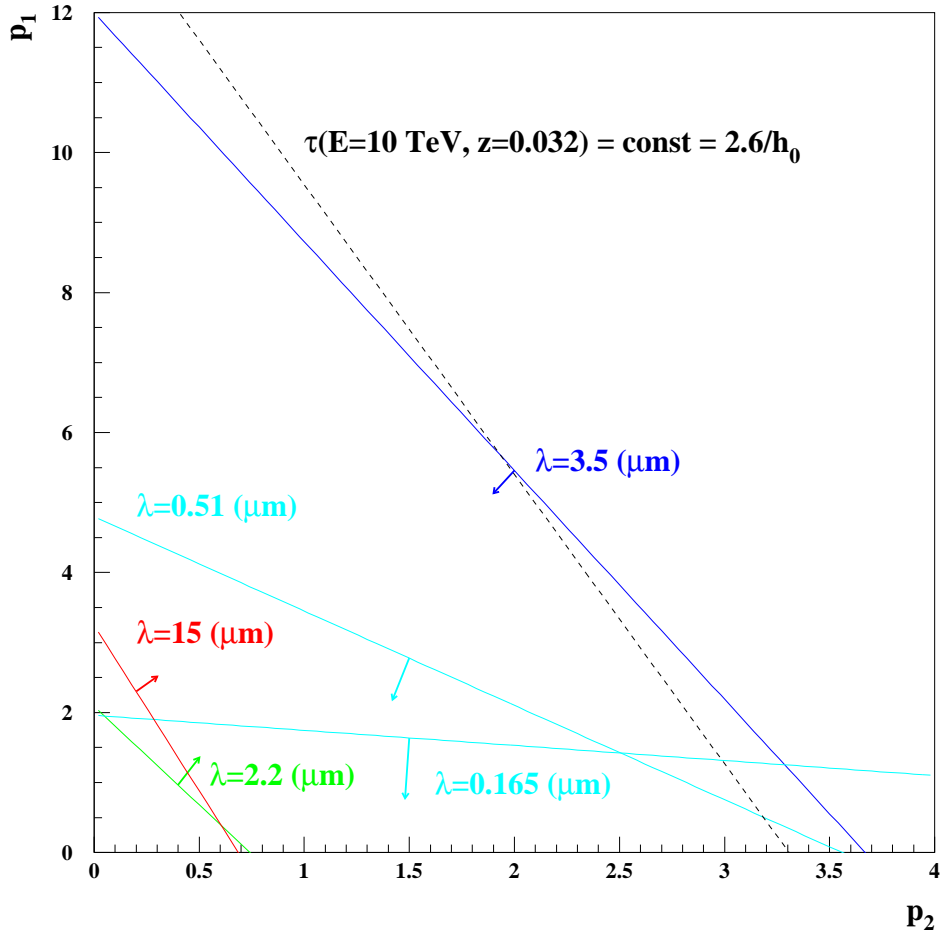,height=5.5in,angle=0.}}
\caption{Constraints on the choice of $p_1$ and $p_2$ parameters 
from existing EBL upper and lower limits. The $p_0$ parameter is 
taken to be $0.025$. The dashed line indicates
constant attenuation of $10$ TeV $\gamma$-rays at $z=0.032$.
The lines parallel to it would correspond to lower or higher 
constant optical depth.}
\label{limits}
\end{figure}

The limitations imposed on the choice of parameters $p_1$ and $p_2$
are shown in Fig.~\ref{limits}. The most severe constraints are due to the
{\em DIRBE} upper limit at $3.5$ $\mu$m and due to lower limits 
found by Pozzetti et al.~\cite{Pozzetti96} for K-band and by Oliver 
et al.~\cite{Oliver97} at $15$ $\mu$m. The upper limits in the wavelength 
range $0.165-0.512$ $\mu$m may potentially constrain the $p_1$ 
parameter as it is shown. At the same time these limits affect little 
the value of $p_2$ which mostly determines the EBL SED at longer wavelengths
and absorption of TeV $\gamma$-rays. The $p_1$ parameter 
may affect the optical depth in the range $0.3-1$ TeV. However, if
survival probability of such photons is close to one, formula
(\ref{EBLsolution}) may not be well justified and in this case 
limits on $p_1$ from the EBL UV region may become invalid.
On the other hand, if the opacity of intergalactic media to 
such $\gamma$-rays is large ($\ge 1$), these limits must be
taken into account. Gamma-ray spectral measurements of sources with 
larger $z$, such as the recently discovered~\cite{Chadwick99} X-ray BL Lac 
PKS 2155-304 $(z=0.117)$, may provide important information
to resolve this question.

In the case of Mrk 421 and Mrk 501, the matrix $t_{ji}(z)$ which 
connects the parameters of the EBL SED to parameters of optical depth is
\begin{eqnarray}
t_{ji}(0.032) &=&
\left( 
\begin{array}{ccc}
   0.043 &   0.007 &  -0.002 \\
   0.080 &   0.041 &   0.003 \\
   0.182 &   0.168 &   0.041 
\end{array}
\right).
\label{z032}
\end{eqnarray}
The elements of the $t_{ji}$ upper triangle are not exactly equal to 
zero because we truncated the SED of EBL at $9.3$ eV, but they are 
negligible. One can see that curvature in the spectra of these AGNs 
is driven by the $p_2$ parameter
\begin{eqnarray}
c_2 & = & \frac{1}{h_0} 0.041 p_2.
\label{c2}
\end{eqnarray}
Although the upper limit on $c_2<0.3$ (see \S\ref{AGNspectra}) is consistent 
with the {\em DIRBE} $3.5$ $\mu$m limit, it provides no additional
constraint on $p_2<7.3 h_0$ for a reasonable range of the normalized
Hubble constant $(0.5<h_0<0.85)$. The preferred value of $c_2=0.17$,
if considered to be dominated by EBL absorption, may give an
estimate of $p_2=4.15 h_0$ which is not ruled out for any $h_0$
from the indicated range. 

The change of the AGN spectral index due to EBL absorption 
\begin{eqnarray}
c_1 & = & \frac{1}{h_0}(0.041 p_1 + 0.168 p_2)
\label{c1}
\end{eqnarray}
can range from $0.12/h_0$ to $0.61/h_0$. This is a large interval since it 
is possible to have $(\sim 0.2-1.0)$ EBL contribution to the $2.28$ spectral 
index of Mrk 501. Currently we are not able to use this EBL constraint.
But once the mechanism of blazar emission is understood and the $c_1$ 
contribution is isolated or limited from the measurements of the AGN spectrum, 
then Eq.\ (\ref{c1}) may become an important restriction on the EBL SED.

   
The total flux from a blazar is another parameter subjected to ambiguous 
interpretation. We find that the optical depth for a $10$ TeV photon
is given by
\begin{eqnarray}
\tau(10\mbox{ TeV},z=0.032) & = & \frac{1}{h_0}(0.05 p_0+0.19 p_1 + 0.79p_2).
\label{od10TeV}
\end{eqnarray}
The line of equal survival probability of such a photon is shown in 
Fig.~\ref{limits}. The maximal and the minimal optical depth
allowed in the limited region of $(p_1,p_2)$ parameters is reached
at $(0.,3.6)$ restricted by the {\em DIRBE} $3.5$ $\mu$m point,
and almost at $(0.,0.7)$ constrained by the Pozzetti et al.~\cite{Pozzetti96}
$2.2$ $\mu$m measurement. Thus the possible range
of optical depths is given by $0.55/h_0 - 2.88/h_0$ or
approximately from $1$ to $5$. In the latter scenario survival 
probability for such a photon is less than 1\%.
Fig.~\ref{SEDlimits} shows the EBL SED for these limiting cases of maximal 
and minimal attenuation of $10$ TeV $\gamma$-rays. This figure 
also depicts an example of the SED of EBL which satisfies a tentative 
EBL detection point reported recently by Dwek \& Arendt~\cite{DwekArendt98}.

Concluding this section we point out an important limitation
which concerns our SED solution for EBL given by Eq.\ (\ref{EBLsolution}).
This limitation arises from the general problem of ``ill-defined'' 
unfolding tasks being very sensitive to small fluctuations 
in the source function~\cite{Tikhonov}, in our case optical depth $\tau(E)$. 
In principle, the EBL unfolding procedure given by Eq.(\ref{Unfold2})
can be performed for arbitrary $\tau(E)$. We, nevertheless, chose a smooth
approximation (\ref{Aopticaldepth}), and obtained a smooth solution
(\ref{EBLsolution}). If $\tau(E)$ is slightly perturbed, 
even with statistically small fluctuations, the unfolded SED of EBL
may acquire large distortions. This fact, mainly originating from the 
asymptotic properties of $F(q)$ (see Eq.~\ref{assymptotic}), 
simply indicates a deficiency in the complete reconstruction of the soft 
photon density by unfolding of the optical depth.
We assume, therefore, that the suggested EBL solution (\ref{EBLsolution})
describes a smooth component of the SED. Small perturbations or spectral 
line features with small amplitudes are possible in the EBL SED 
as long as they are not detectable in $\tau(E)$ because 
integration (\ref{Unfold1}) smoothes them out drastically. 
We also note that if the EBL SED is very small, close to the indicated 
lowest limit, then formula (\ref{EBLsolution}) may also become
inapplicable, except in the high energy end, because EBL 
absorption will be virtually undetectable in the rest
of the spectrum.

\section{Attenuation of $\gamma$-rays by EBL}
\label{EBLattenuation}

An uncertainty in the interpretation of AGN spectral measurements 
when determining an attenuation of the AGN absolute flux 
and change of its spectral index because of the EBL intergalactic
absorption creates difficulty in the prediction of the visibility 
range of TeV blazars. We use our maximal and minimal 
estimates of the EBL SED, found in the previous section, to predict 
the maximum and the minimum effect of EBL on the observable spectrum of 
a ``Mrk 501-like'' blazar at higher redshifts if such a hypothetical 
object demonstrates a peak of activity. Then we compare its flux with the 
sensitivities of the existing Whipple~\cite{Weekes89} and proposed 
VERITAS~\cite{VERITAS} $\gamma$-ray observatories.

\begin{figure}
\centerline{\epsfig{file=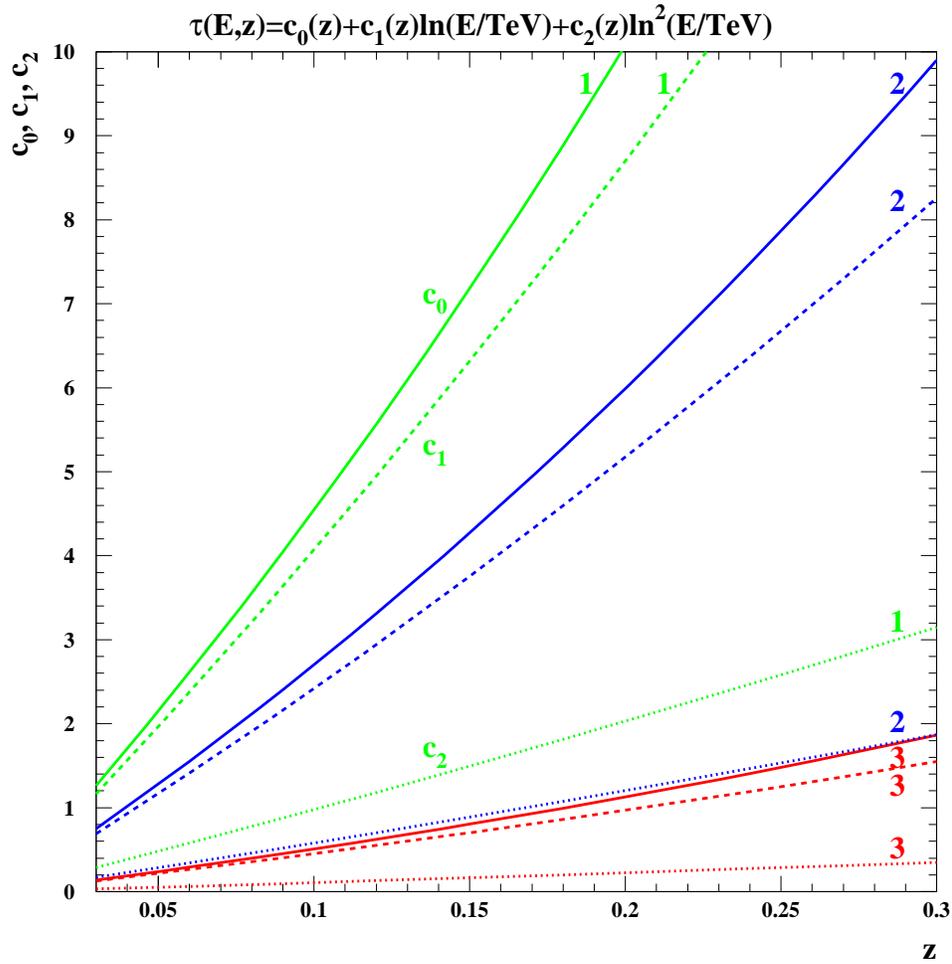,height=5.5in,angle=0.}}
\caption{The dependence of $(c_0(z),c_1(z),c_2(z))$ functions
on redshift for the highest allowable EBL absorption effect (marked $1$),
for the lowest already detected EBL density (marked $3$), and for the
probable EBL density (marked $2$, see explanation in text).
The $c_0$ curves are shown as solid lines. The $c_1$ and $c_2$ 
curves are shown as dashed and dotted lines respectively.}
\label{c0c1c2}
\end{figure}

We extrapolate the optical depth $\tau(E,z)$ to higher redshifts 
by calculating the $t_{ji}(z)$ matrix. Fig.~\ref{c0c1c2} demonstrates 
the dependence of $c_0(z),c_1(z),c_2(z)$ of the 
parameterized $\tau(E,z)$ expression (Eq.~\ref{Aopticaldepth}).
This figure shows three sets of curves marked $1$, $2$, and $3$.
The first set corresponds to a maximal allowed absorption. The third
describes the minimal EBL effect based on the already resolved EBL SED
from galaxy counts in deep surveys of extragalactic sources.
The second set of curves indicates the SED of EBL defined by
$(p_1,p_2,p_3)=(0.025,0.,4.15 h_0)$, which is preferable if 
curvature in the spectra of Mrk 501 and Mrk 421 is dominated 
by EBL absorption (see \S\ref{AGNspectra}).

\begin{figure}
\centerline{\epsfig{file=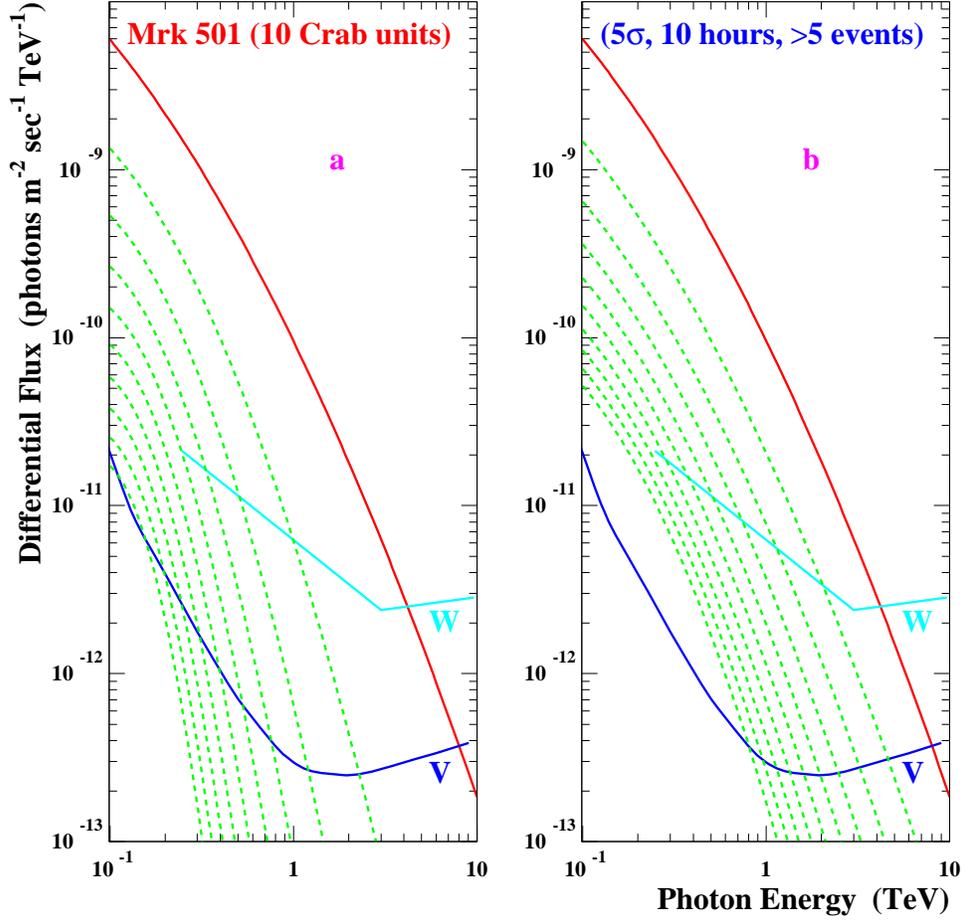,height=5.5in,angle=0.}}
\caption{The change of the observable spectrum of a hypothetical
``Mrk 501-like'' object during its highest activity if
it is placed at higher redshifts. Figure ``a'' shows the maximum
allowable EBL absorption effect. Figure ``b'' indicates the 
opposite limit. The sensitivities of Whipple and VERITAS are
marked as ``W'' and ``V'' (see explanation in text).
The solid line shows the Mrk 501 spectrum $(z=0.03)$ extrapolated 
into the $100-250$ GeV region. The absolute flux of the blazar is 
$10$ Crab units. The dashed lines correspond to the spectra
of such AGN if observed at redshift $z=0.06, 0.09, ...,0.3$.}
\label{Mrk501attenuation}
\end{figure}

Figure~\ref{Mrk501attenuation} summarizes changes in the  
spectrum of a ``Mrk 501-like'' blazar placed at 
redshifts $z=0.03, 0.06,..., 0.3$. The total flux of this
hypothetical AGN is normalized to $10$ Crab units, the 
highest flux ever observed from Mrk 421~\cite{Gaidos96}. The sensitivities of 
Whipple and VERITAS, marked ``W'' and ``V'', are indicated~\cite{VERITAS}. 
They are derived for certain cuts on parameters of the atmospheric 
shower images and correspond to a $5\sigma$ detection in $10$ hours, 
or detection of at least five photons in the regime of low 
background. The sensitivity shown reflects the standard, small zenith angle,
observation mode. In the high energy end it can be substantially 
improved by utilizing the large zenith angle observation 
technique~\cite{Krennrich98}. Two plots ``a'' and ``b'' demonstrate 
the absorption effect for the highest allowable and the lowest resolved 
EBL. The change of the flux at $100$ GeV when $z$ 
increases is mostly due to a simple distance factor $\sim z^{-2}$.
The intergalactic absorption of such $\gamma$-rays is small
relative to this effect, and our EBL attenuation prediction, which is
not reliable in this energy regime, is masked by this factor.
For the lowest EBL SED, the distance effect dominates absorption for 
all shown curves in the energy interval below $2$ TeV.
In this case, the Whipple telescope should be able to detect such 
a blazar up to $z\sim 0.18$ during a few days long
episode of activity. For the worst-case scenario of intergalactic 
absorption the Whipple visibility range to such a hypothetical
object is limited to $z\sim 0.12$. Thus, it seems that the high variability 
of blazars is the reason which so far prevents us from detection 
of such more distant objects, or luminosity of potential TeV blazars
is too low to identify them at high redshifts as X-ray emitters. The recently 
detected X-ray selected BL Lac PKS 2155-304 has been observed to have
maximal X-ray output in the range $0.05-0.5$ keV during active 
states~\cite{Urry97}, which is similar to Mrk 421 peaking
at a few keV~\cite{Buckley97}. This object showed a similar
observed luminosity in X-ray but it is located at
$z=0.117$. The detection of such a more luminous but more
distant BL Lac is certainly not limited to $z<0.12$ and 
may extend to higher redshifts.

The new generation of $\gamma$-ray observatories,  
such as VERITAS~\cite{VERITAS} and HESS~\cite{HESS}, 
should be able to detect ``Mrk 501-like'' blazars at least 
up to distances of $z=0.3$ as shown in Fig.~\ref{Mrk501attenuation}a. 
Due to their much improved ability to survey the sky, the list
of identified TeV and sub-TeV emitters will substantially
increase. It is interesting to denote then, that these proposed projects
may indeed {\em measure} the EBL because of the extension of their 
energy range to below $100$ GeV. The SED of the EBL is expected to
drop very rapidly in the UV above $2$ eV making the universe
mostly transparent to $\gamma$-rays with sub-100 GeV energy. 
Therefore, in the vicinity of $100$ GeV one should 
see a ``knee-like'' transition region in the spectra of AGNs
where one power-law is replaced with one 
with a larger spectral index. The position of this feature should 
be independent of characteristics of blazars, and the accumulated change 
of spectral index and change of flux amplitude for several objects 
at known redshifts will define the EBL SED as shown in our calculations. 
The more accurate EBL unfolding will be possible because measurements 
in the region 100 GeV and below provide a reference point in the 
spectra where no EBL absorption effect should be present. 
The remaining uncertainty related to a particular source intrinsic spectrum,
perhaps, will be eliminated when more sources are detected. The visibility 
range of VERITAS and HESS will not be limited to $z<0.3$ if TeV and 
sub-TeV emitters can have a much higher luminosity than Mrk 501.

\section{Discussion}
\label{Discussion}

The {\em DIRBE} collaboration has completed analysis of the 
data on direct measurements of the EBL spectral density
and published detections and upper limits~\cite{Hauser98},
which in many wavelength bands substantially exceed limits
interesting for the theoretical understanding of the  
history of the universe. Currently, only TeV $\gamma$-ray astronomy can 
potentially produce any additional constraints in this field.
At the moment though, its role as a contributor to our knowledge
about EBL was characterized by the {\em DIRBE} collaboration 
in a following conclusion~\cite{Dwek98}. 
``Currently, there is no evidence for any intergalactic absorption 
in the spectrum of the three TeV sources detected to date. 
Claimed detections of, or upper limits on the EBL from TeV
$\gamma$-ray observations should therefore be regarded as 
preliminary at present.'' We acknowledge the fact that 
some papers seeking to use TeV measurements to detect or limit 
EBL in some cases may have been based on preliminary
judgments of unverified results. However, reliable conclusions
derived from TeV $\gamma$-ray observations do exist.

\begin{figure}
\centerline{\epsfig{file=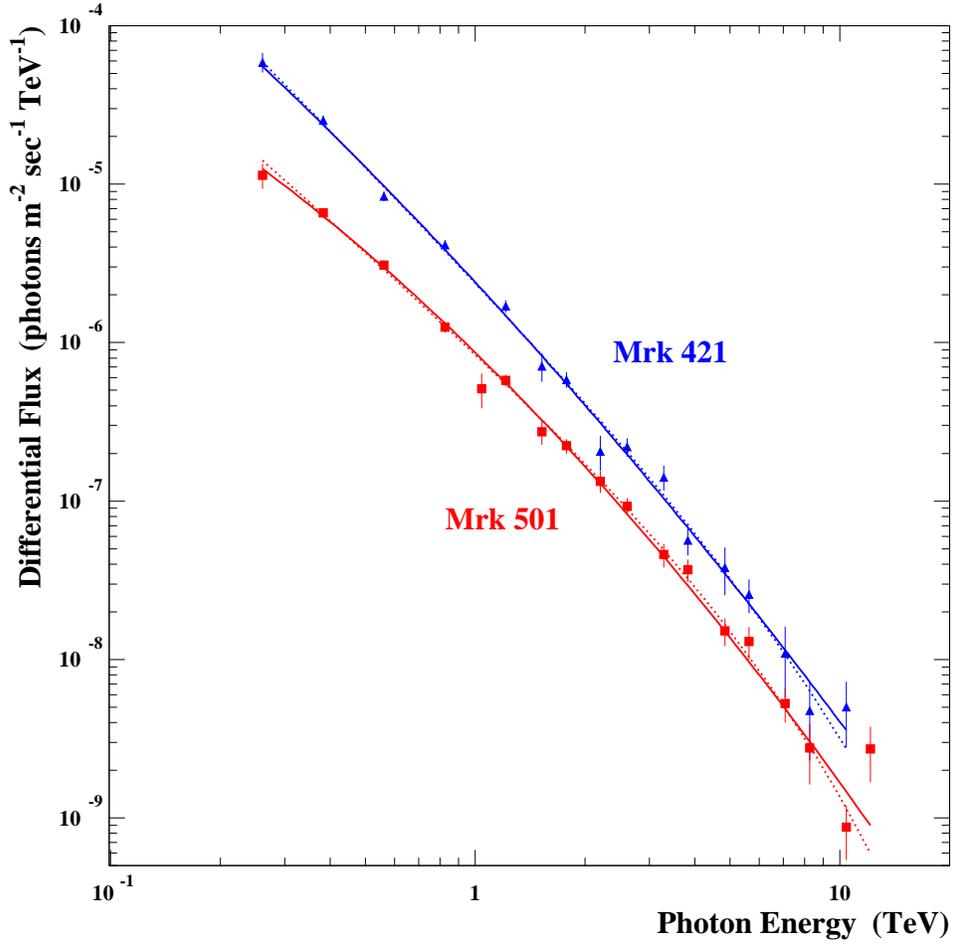,height=5.5in,angle=0.}}
\caption{Spectra of Mrk 421 (filled triangles) 
and Mrk 501 (filled squares) as reported in~\cite{Krennrich98}. 
The solid lines indicate the best fits of the curved spectra 
$\sim \exp\left(-\delta \ln(E/\mbox{TeV}) - c  \ln^2(E/\mbox{TeV}) \right)$ 
with parameters $(\delta, c, p_{\chi^2})$ for Mrk 421 
$(2.50,0.12,0.14)$, and Mrk 501 $(2.26,0.20,0.31)$. 
The dotted lines show the best fits of the spectra with exponential cutoff 
$\sim \exp\left(-\delta \ln(E/E_c) - E/E_c \right)$ and parameters
$(\delta, E_c, p_{\chi^2})$ for Mrk 421 $(2.32,7.0\mbox{ TeV},0.11)$,
and Mrk 501 $(1.99,4.9\mbox{ TeV},0.41)$. Due to variability of the 
sources the absolute fluxes shown have arbitrary normalization.}
\label{Mrk421Mrk501}
\end{figure}

\begin{figure}
\centerline{\epsfig{file=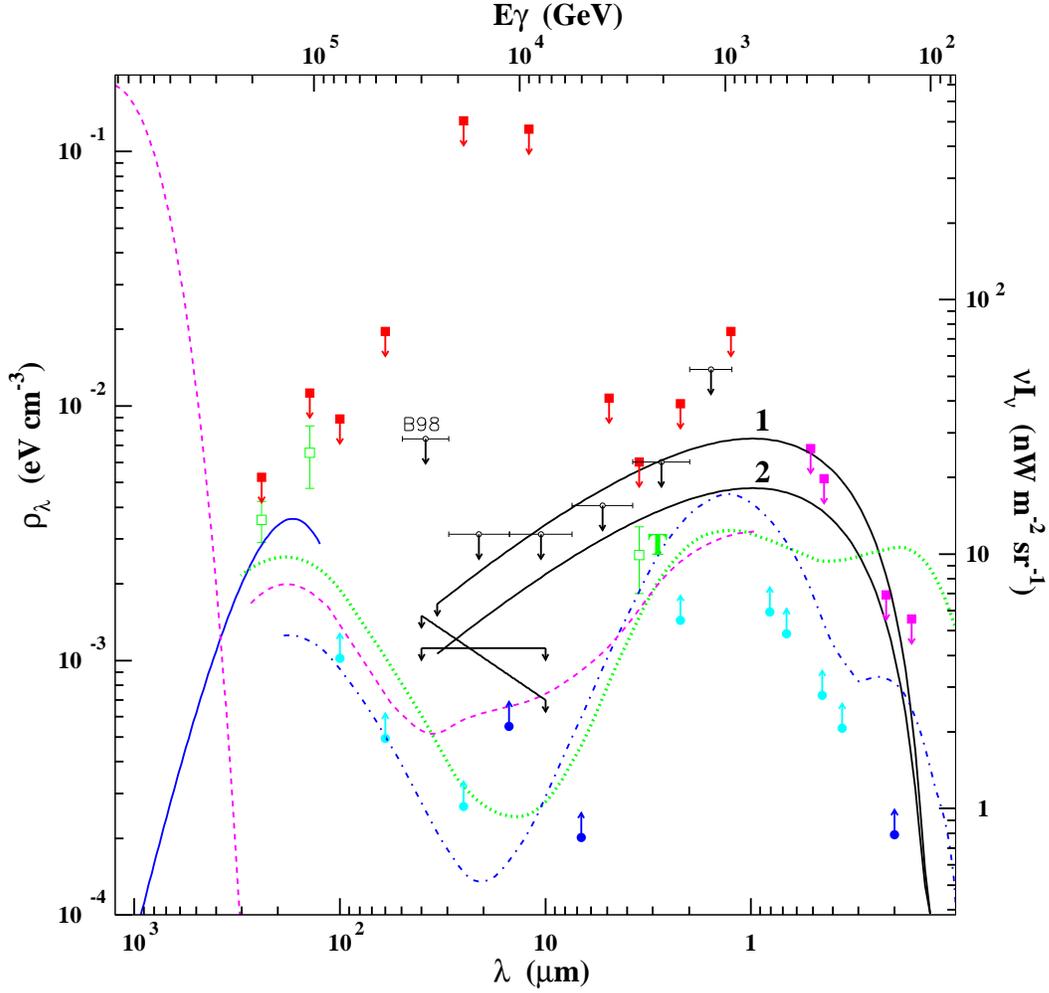,height=5.5in,angle=0.}}
\caption{The solid line marked ``1'' provides an upper limit in the 
range $\sim 1-30$ $\mu$m on the SED EBL derived in this work.
The range of the EBL SED between curves ``1'' and ``2'' is 
preferable if all curvature observed in the spectra of Mrk 501 
and Mrk 421 is attributed to EBL absorption.  
Upper limits from the Mrk 501 TeV spectrum by Biller, 
et al.~\cite{Biller98} normalized at the $3.5$ $\mu$m {\em DIRBE} upper limit  
are shown as horizontal bars with arrows.
``No cutoff'' upper limits from Mrk 421 spectra are indicated in the range 
$10-40$ $\mu$m by solid line segments with arrows (see explanation in text). 
Dotted line is LCDM (Salpeter) model by Primack et al.~\cite{Primack99},
and dot-dashed line is LCDM (Scalo) model from the same publication.
Dashed line is ``high'' EBL model by Malkan \& Stecker~\cite{MalkanStecker98}.
Filled squares and circles indicate upper and low EBL limits respectively 
(see Fig.~\ref{EBLdata} for explanation).}
\label{summary}
\end{figure}

The ``no cutoff'' approach to derive upper limits on EBL discussed 
in \S\ref{EBLinteraction} is a valid method as long as assumptions 
of the model are justified in the relevant wavelength interval.
It might be arguable if the original calculations, presented in the papers 
\cite{SDS92, SD93, DeJager94}, have always satisfied this requirement. 
According to the current paradigm, though, the wavelength regime 
$10-40$ $\mu$m is expected to be dominated by emission from dust, 
for which EBL spectral density (\ref{spdn}) with $\beta=2.0-3.0$ 
provides an acceptable description. The absence of a cutoff
in the spectrum of Mrk 421 at least up to $E_c=8.8$ TeV ($\beta=2.55$),
or $E_c=7.0$ TeV ($\beta=2.00$) translates to the upper limits on the SED 
of EBL (see Eqs.\ (\ref{EBLcutoffL} \ref{r255}, \ref{r200}) and 
Fig.~\ref{Mrk421Mrk501})
\begin{eqnarray}
\rho_{\lambda} < & 1.6 \times 10^{-3} h_0 & 
\mbox{ eV cm$^{-3}$    ,    } 2.0\ \mbox{$\mu$m} <\lambda<  37\ \mbox{$\mu$m} 
\label{EBL200} \\
\rho_{\lambda} < & 1.0 \times 10^{-3} h_0
\left(\frac{\lambda}{\mbox{10 $\mu$m}}\right)^{0.55} & 
\mbox{ eV cm$^{-3}$    ,    } 4.1\ \mbox{$\mu$m} <\lambda<  40\ \mbox{$\mu$m}. 
\label{EBL255}
\end{eqnarray}
Unless the intrinsic spectra of both Mrk 421 and Mrk 501 are exponentially 
rising to exactly compensate EBL absorption, these limits constrain rather 
strongly existing EBL models in this wavelength range (see Fig.~\ref{summary}) 
as they are almost two orders of magnitude lower than the {\em DIRBE} upper 
limits at $12$ and $25$ $\mu$m.  Kashlinsky et al.~\cite{Kashlinsky96a, 
Kashlinsky96b} derived comparable limits from the search for a fluctuation 
signal in the {\em DIRBE} maps which reflects the spatial correlation 
properties of the galaxies implying that EBL comes from the redshifted light 
emitted by them. This method relies upon an extended set of assumptions 
on the evolution of such parameters as the galaxy SED, galaxy luminosity 
function with redshift, galaxy clustering, dust absorption and re-emission.
The effects of intervening dust, among these, are less understood,
especially for this wavelength range because the EBL re-processing 
by dust could have occurred outside of the galaxy which produced it.
The limits derived from the ``no cutoff'' observations of 
$\gamma$-rays, on the other hand, require only the assumption of 
the EBL SED behavior in a certain wavelength interval and they are
applicable {\em only} to models which satisfy this condition.
If one considers the other limits for this wavelength range 
(shown in Fig.~\ref{summary}) derived from the different assumptions 
explained below, one can conservatively limit the SED in $10-40$ $\mu$m
interval to be below $10$ nW m$^{-2}$ sr$^{-1}$.


No apparent cutoff due to intergalactic absorption 
has been observed in the $0.26-10$ TeV spectra 
of the two most-studied AGNs. However, the recently reported 
spectra of Mrk 421 and Mrk 501~\cite{Krennrich98}
are not inconsistent statistically with a common major 
contributor to the curvature of the spectra of these 
blazars, perhaps EBL absorption. This explanation 
might be preferable because almost pure intrinsic power-law 
of these sources provides a larger degree of invariance 
to support the lack of spectral shape variability observed
during flaring activity in Mrk 501. This result, however,
needs further confirmation. 

The fact that no distinct cutoff feature has been 
observed in the spectra of two blazars allows a non-trivial
interpretation of the EBL SED, as shown in this paper. 
The spectral energy density should have a certain form 
and coincide with the expected peak in this wavelength 
region of the SED from starlight emitted and
redshifted through the evolution of the universe. It is rather a 
peculiar coincidence that TeV $\gamma$-ray observations 
currently span an energy range in which the contribution from the
EBL absorption to the spectra of extragalactic sources can be 
masked easily by unknown properties of the source which has
a spectrum close to a pure power-law. Extending the energy 
range to either higher energies or lower ones should 
identify a cutoff and knee-type features generated by the 
dust and stellar components of EBL, respectively.

At the moment the upper limit on the EBL SED can be found 
(see Fig.~\ref{summary}) based on two assumptions.
First, we normalize the EBL spectral density at the $3.5$ $\mu$m 
{\em DIRBE} upper limit and, second, we propagate the SED solution 
to shorter and longer wavelengths requiring that such an EBL SED 
may generate changes only in the source flux amplitude, 
spectral index of the power-law, and spectral curvature.   
These are allowed by observations of Mrk 501 and Mrk 421.
The expected rapid decline of the EBL SED in the UV region is
also used in these considerations. Our limits 
in the range $1-10$ $\mu$m are consistent with those previously 
derived by Biller et al.~\cite{Biller98} and by 
Dwek \& Slavin~\cite{DwekSlavin94}. The former work also
used the {\em DIRBE} upper limit for normalization
and propagated the allowed SED solution by limiting the range of 
spectral index changes. Dwek \& Slavin were first to
study $\gamma$-ray absorption by the stellar component 
of EBL and found high model dependent upper limits 
in this wavelength range. This is the first work 
which indirectly proved that a substantial amount of the EBL
produced by redshifted starlight may escape detection 
in TeV $\gamma$-ray observations of a source with unknown 
absolute flux and spectral index. 

The limits derived from TeV $\gamma$-ray measurements in $1-10$ $\mu$m region 
seem to be in good agreement with each other except the very 
low ones given by Stanev \& Franceschini~\cite{StanevFranceschini98}.  
These authors also attempted to unfold EBL spectra indirectly
from the measurements of Mrk 501, but the absolute flux normalization and 
power-law spectral index were allowed to vary during the fitting procedure. 
As we showed in \S\ref{EBLunfolding}, an EBL SED component of the form  
$\varepsilon(p_0-p_1 \ln (\varepsilon/\varepsilon_0))$
with arbitrary constants is likely to be undetectable 
with such a method. Perhaps existence of this hidden 
invariance was the reason for unstable behavior of the
solution which has been found to be crucially dependent on 
the manner in which power-law fluxes are normalized.  
If our supposition is valid such a component can be added 
to the very low upper limits given in this paper until
it becomes constrained by one of the experimental points,
likely the {\em DIRBE} $3.5$ $\mu$m upper limit.

Our low and high estimate of the opacity of the universe 
to TeV and sub-TeV $\gamma$-rays indicate that attenuation
should not prevent us from seeing more TeV
sources even with existing $\gamma$-ray observational
facilities. The future projects, such as VERITAS, HESS, and MAGIC,
have a high potential to increase the list of TeV and 
sub-TeV sources, because they extend their sensitivity 
range for objects similar to the TeV emitter Mrk 501 
up to $z=0.3$. If there are more luminous objects capable
of producing TeV radiation, this redshift will not be a limit.

The TeV and sub-TeV observations have the potential
to further contribute in identifying the EBL SED. The
HEGRA collaboration has measured recently the Mrk 501
spectrum with very high statistical quality~\cite{Aharonian99a}.
For the first time the energy range of detected $\gamma$-rays
is extended above $10$ TeV up to $\sim20$ TeV providing
a sensitivity in the $40-70$ $\mu$m region of EBL. 
The rapid decline in the Mrk 501 spectrum found at the highest 
observed energies suggests several explanations such as
intergalactic EBL absorption due to the dust component, intrinsic 
absorption of $\gamma$-rays at the source, or a cutoff in the 
photon production mechanism~\cite{Coppi99, Konopelko99a}.
Further progress in the analysis of this data
can be achieved if spectral and temporal characteristics 
of the emission in X-ray and TeV regions are found during 
multiwavelength campaigns or a cutoff feature is identified
in the spectrum of Mrk 421 above 10 TeV. In both cases, this will
reduce the interpretation ambiguity and perhaps the dust EBL component 
expected in this wavelength interval will be detected or
strongly constrained.

\begin{figure}
\centerline{\epsfig{file=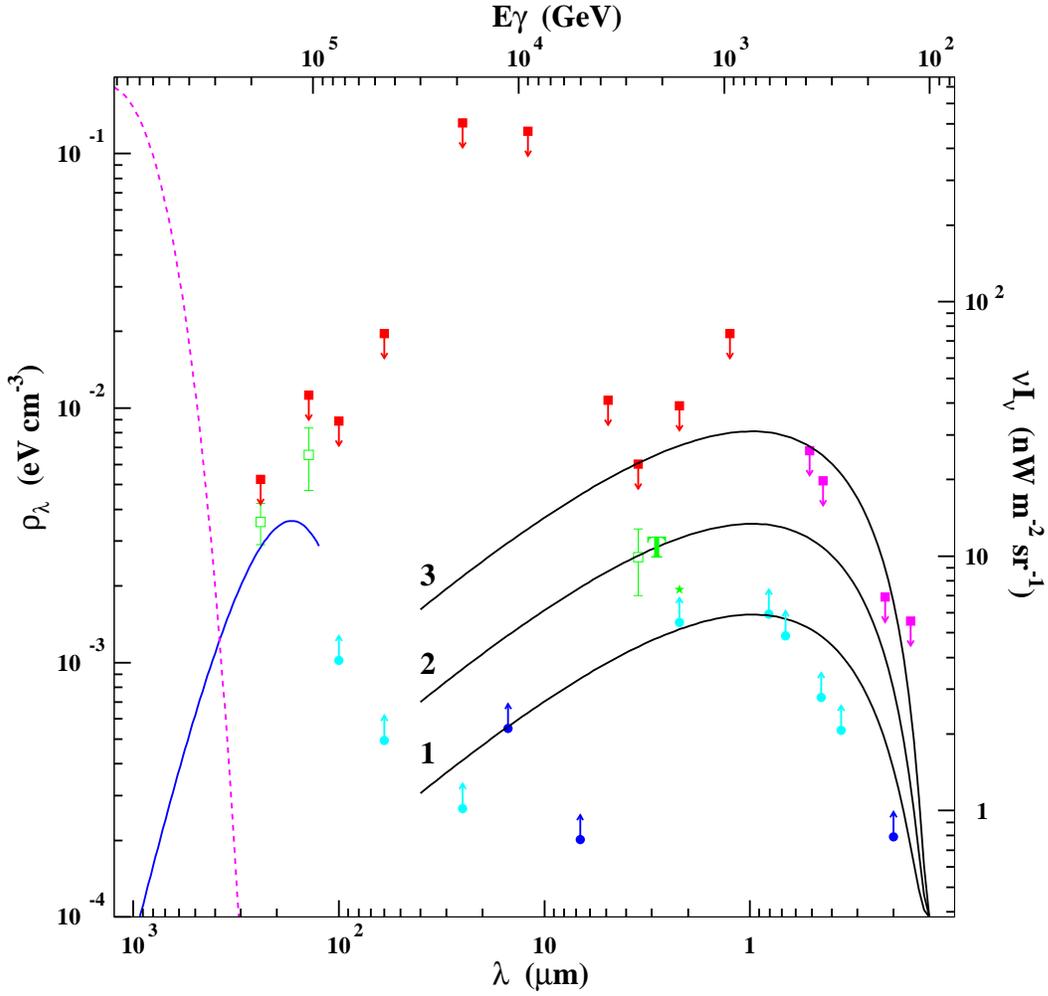,height=5.5in,angle=0.}}
\caption{The examples of SED EBL corresponding to maximal and minimal
attenuation of $10$ TeV $\gamma$-rays consistent with EBL upper and low
limits and with observations of the Mrk 421 and Mrk 501 TeV spectra.
Line $1$ defined by $(p_1,p_2,p_3)=(0.025,0.,0.7)$ gives minimal
optical depth $0.55/h_0$. Line $3$, $(0.025,0.,3.6)$ generates 
maximal allowable optical depth $2.88/h_0$. Line 2, $(0.025,0.,1.6)$,
shows an example of SED EBL which is consistent with recent tentative
EBL detection at $3.5$ $\mu$m~\cite{DwekArendt98}. This EBL distribution
would be responsible for $1.26/h_0$ optical depth.}
\label{SEDlimits}
\end{figure}

The valley between $10-40$ $\mu$m  may also become testable. 
The empirically based calculations of the EBL spectrum by 
Malkan \& Stecker~\cite{MalkanStecker98} produce a fairly flat 
EBL SED in this region (see Fig.\ref{summary}). This feature 
of their EBL model is primarily responsible for a prediction 
by Stecker~\cite{Stecker99} of an absorption cutoff above 
$\sim 6$ TeV for PKS 2155-304 $(z=0.117)$. It may be
that a very large exposure on the source may be necessary
to deduce spectrum up to this energy, but if this cutoff is not
observed, it will indicate a SED in this region
closer to the lines ``1-3'' shown in Fig.~\ref{SEDlimits}
rather than an almost constant EBL SED distribution.

The limits in the wavelength region $\sim 1-10$ $\mu$m 
can be improved if the EBL signal in the spectra of AGNs 
is isolated, or at least limited, by blazar emission models.
Currently, an upper limit on the spectral curvature of 
Mrk 501 and Mrk 421 does not provide a more strict limit 
than the {\em DIRBE} direct upper limit at $3.5$ $\mu$m.
However, if the most likely curvature, $0.17$, in their spectra is 
due to EBL absorption then the range of possible SEDs is limited
by lines marked as ``1'' and ``2'' in Fig.~\ref{summary}.
These predictions seem to be higher than current
EBL theoretical models. This could mean either an additional
contribution to EBL from an unidentified source, or that at least 
part of the blazars' spectral curvature must be intrinsic to the sources. 
If the latter is correct, then confirmation of spectral invariance 
during episodes of activity of blazars may strongly limit flaring
mechanisms. Restricting the change of spectral index, $c_1$, due to EBL 
absorption may impose strict limits on the SED in $\sim 1-10$ $\mu$m 
range. Measurements in the GeV range are available for 
Mrk 421~\cite{Hartman99} which show spectral index $1.57$.
The spectral index for the best fit in the $0.26-10$ TeV range
with curvature $0.17$ gives value $2.48$. The $0.91$ 
difference again provide a rather loose constraint on EBL
since we already limited this number to be between $0.12/h_0$
and $0.61/h_0$. It is possible to explain such a change in the blazar
spectral index purely by $\gamma$-ray absorption, again
implying a substantial SED of EBL limited by lines ``1'' and 
``2'' depending on the value of $h_0$. There is no argument, however,
to support that such change in spectral index could not be 
intrinsic to the source since these measurements are 
separated by more than two orders of magnitude in energy. 
The recent report of a $5.2\sigma$ ($E>0.5$ GeV) detection 
of Mrk 501 by EGRET~\cite{Kataoka99} estimated the spectral index
as $1.3\pm 0.5$ again indicating that the largest allowable
opacity to TeV photons is possible. It seems that 
constraints on EBL in the $1-10$ $\mu$m wavelength region may 
come only if the spectrum of a new TeV source will be measured,
preferably in TeV, sub-TeV, and GeV regions.

The EBL region $0.1-1$ $\mu$m is not constrained well 
by current sub-TeV observations. This wavelength interval 
holds crucial information which may finally allow one
to determine EBL through the whole range $\sim 0.1-10$ $\mu$m.
VERITAS, HESS and MAGIC  should be able to see farther and 
detect many sources. It is very important that the energy range 
of these observatories extend to below $100$ GeV 
providing a possibility to measure in one experiment the
``turn on''of the effect of sub-TeV $\gamma$-ray absorption 
by EBL. The ``knee-like'' feature has to exist in the spectra
of blazars around $100$ GeV, independent of properties of AGNs. 
The change of the power law through this knee, if observed for several
sources, will determine the amount of EBL in this wavelength band. 
It is very important also that due to the high
sensitivity of these experiments the spectra of the closest
AGNs Mrk 421 and Mrk 501 will perhaps be measured up to $~50$ 
TeV, providing information on the $70-140$ $\mu$m EBL window.
This will close the last gap in the EBL SED and meet with direct 
detections by the {\em FIRAS} and {\em DIRBE} experiments.

\begin{ack}
I thank F. Krennrich of the Whipple collaboration for providing the spectra 
of Mrk 421 and Mrk 501, J. Primack and J. Bullock for useful discussion, 
M. Catanese for valuable discussions and invaluable help, and T. Weekes 
for thoughtful advice and encouraging this study. This work was supported 
by grants from the U.S. Department of Energy.
\end{ack}


\newpage
\pagestyle{empty}

\newpage

\newpage

\newpage

\newpage

\newpage

\newpage

\newpage

\newpage

\newpage

\end{document}